\documentclass{article}

\usepackage{arxiv}
\usepackage[utf8]{inputenc}
\usepackage[T1]{fontenc}
\usepackage[english]{babel}
\usepackage{ifpdf} 
\usepackage{newtxtext} 
\usepackage{newtxmath} 
\usepackage{array,graphicx,dcolumn,multirow,abstract,hanging,fancyhdr,float}

\usepackage{amsthm}
\usepackage{tikz}
\usepackage{tikz-qtree}
\usetikzlibrary{trees,calc,arrows.meta,positioning,decorations.pathreplacing,bending}
\tikzset{
    edge from parent/.style={draw, thick, blue!70!black},
    no edge from this parent/.style={
        every child/.append style={
        edge from parent/.style={draw=none}}},
    level 3/.style={yshift=5cm},
    level 4/.style={level distance=5mm} 
         }
\usepackage[linesnumbered,ruled,vlined]{algorithm2e}
\usepackage[labelfont=sc,textfont=sf]{caption}
\usetikzlibrary{shapes.misc}
\newcommand*\cancel[2][thin]{\tikz[baseline] \node [strike out,draw,anchor=text,inner sep=0pt,text=black,#1]{#2};}

\newcolumntype{L}[1]{>{\raggedright\arraybackslash }p{#1}} 
\newcolumntype{C}[1]{>{\centering\arraybackslash }p{#1}}
\newcolumntype{R}[1]{>{\raggedleft\arraybackslash }p{#1}}
\newcolumntype{d}[1]{D{.}{.}{#1}} 

\let\tempone\itemize
\let\temptwo\enditemize
\let\tempthree\enumerate
\let\tempfour\endenumerate
\renewenvironment{itemize}{\tempone\setlength{\itemsep}{0pt}}{\temptwo}
\renewenvironment{enumerate}{\tempthree\setlength{\itemsep}{0pt}}{\tempfour}

\newtheorem{theorem}{Theorem}

\newtheorem{claim}[theorem]{Claim}

\newtheorem{lemma}[theorem]{Lemma}

\newtheorem{proposition}[theorem]{Proposition}

\newcommand{\mldiamond}{\lozenge}
\newcommand{\smldiamond}{\blacklozenge}

\title{On link deletion and point deletion in games on graphs}

\author{
 Sujata Ghosh \\
 Indian Statistical Institute, Chennai, India\\
 sujata@isichennai.res.in\\

   \And
Shreyas Gupta \\
Indian Institute of Science, Bangalore, India\\
shreyas17521@iisc.ac.in\\
  \And
Lei Li\\
 Tsinghua  University, China \\
 lilei19@mails.tsinghua.edu.cn\\

}

\begin{document}
\maketitle
\begin{abstract}
{\footnotesize We discuss link and point deletion operators on graph games and provide a comparative logic-algorithmic study of the same. In  particular, we focus on 
a popular notion of invariance in transition systems, namely, bisimulation, between the respective games on graphs. We present both logical and 
algorithmic analyses of the concepts so as to provide a more formal analysis of the natural connection between these two operators.}
\end{abstract}
\section{Introduction}

\small

In two-player games on graphs we generally consider two players playing a turn-based game by moving a token through a directed graph, tracing out a finite or infinite path.
Such games provide us with a powerful tool to reason about various question arising in diverse domains, e.g., computer science, logic, linguistics, economics, mathematics, 
philosophy, and biology. One can also consider different variants of such graph games where such variations can arise from different winning conditions (e.g., reachability, parity \cite{graedel11}),
independent moves of players (e.g., cop and robber game \cite{cop-robber}), one player obstructing moves of the others (e.g., sabotage game \cite{sabotage}, poison game \cite{poison}) and others. 
In the interplay between game theory, logic and computer science, these graph games provide good models for reactive systems that need to interact with the uncertain environment.

We now focus on one such variant mentioned above, namely, graph games where one player obstructs the moves of the other player by bringing in some structural changes in the underlying graph.
From the perspective of link/edge deletion in graphs, sabotage games \cite{sabotage} are natural examples where one player is concerned with a reachability objective and the other player is involved in obstructing 
her opponent's moves by deleting edges from the graph. In \cite{sabotage2}, the study has been extended to weighted graphs with multiple destinations with both local and global link deletions. 
Algorithmic studies on local link deletion can be found in \cite{zhang}.

A game that is close to the spirit of games describing point/vertex deletion on graphs is the poison game \cite{poison}: One player (mover) is concerned with moving in the game graph, 
and the other player is involved in obstructing 
her opponent's moves by poisoning certain vertices whose effect is analogous to that of `point deletion' from the  mover's perspective. One can also consider variants of these poison games, 
termed as occupation games \cite{vbl2020}, where the mover 
might also have a reachability objective. We note here that such games, where both players delete links or points or both, occur abundantly in the combinatorial game theory 
literature (e.g., see \cite{vertex1,vertex2,vertex3}). Thus, studying these
games using logic and algorithms provides us with various tools for modelling interactive phenomena in different domains.

Sabotage modal logic (SML) \cite{sab-logic} provides a natural language to reason about link deletion in graphs. Modal logic of stepwise removal (MLSR) \cite{point-logic} provides a natural language to reason 
about point deletion in graphs. So, in the remaining part of the paper, we focus on these two logics and their variants (cf. Section 2) for our bisimulation study on the game graphs. 
Complete proof systems for SML and MLSR have been discussed in \cite{sab-logic,point-logic}, respectively. For the decidability and complexity questions, we have the following results: (i) The satisfaction problems for SML \cite{fsttcs2003} 
and MLSR \cite{point-logic}  are both undecidable, and (ii) the model-checking problems for SML \cite{mfcs2003} and MLSR \cite{point-logic} are both PSPACE-complete. A result that is missing in this picture is the complexity of bisimulation or the model comparison problem,
and in this work we investigate this issue to provide a better understanding of the model comparisons in the respective logics and their inter-relationship. To the best of our knowledge, this study of bisimulation is the last major open complexity problem concerning these graph logics of link or point deletion. Solving this problem will, on one hand, provide us with a finer understanding of the practical applicabilities 
of these logics, and on the other hand, provide us with better insights about their expressive powers. 

The rest of the paper can be summarized as follows: In section 2, we introduce the relevant logic frameworks together with their respective notions of bisimulations. Section 3 deals with logical characterizations of the notions of bisimulation in terms of model-checking in their respective languages, suggesting upper bounds for the problems. Section 4 gives us a detailed algorithmic and complexity-theoretic study of these concepts, validating the upper bound suggestions of section 3. Section 5 provides some further related results and concludes the paper with a discussion on the lower bound.

\section{On link deletion and point deletion logics}

For the sake of completeness, we now provide a discussion of the relevant logics of link deletion and point deletion. We start with a brief outline of 
sabotage modal logics followed by modal logics of stepwise removal.

\subsection{Sabotage modal logics}

We first provide the language and semantics of a generalized version of SML (GSML), similar to what is proposed in \cite{lilei}. 
Given a countable set of propositional variables $\mathcal{P}$, the formulas of GSML are given as follows:

 \begin{quote}
       $\varphi\coloneqq p\mid\neg\varphi\mid(\varphi\land\varphi)\mid\mldiamond\varphi\mid\smldiamond^{\varphi}_{\varphi}\varphi$,
    \end{quote}
    
    \noindent  where $p\in\mathcal{P}$. The intuitive meaning of the formula $\smldiamond^{\psi}_{\chi}\varphi$ is as follows: after some edge is deleted 
 from the model whose end-points satisfy $\psi$ and $\chi$, respectively, the formula $\varphi$ still holds at the point of evaluation. The models for GSML are given by  standard relational models $\mathcal{M} = (W, R, V)$ for modal logics,
where, $W$ is a non-empty set, $R \subseteq W \times W$, and $V : \mathcal{P} \rightarrow 2^W$. A pair $(\mathcal{M}, w)$, 
where $w \in W$ is called a \emph{pointed model}. The truth definition 
of GSML formulas in such pointed models given by $((W,R,V), w)$ are as usual for the propositional, boolean and the modal formulas, and for the new sabotage modality it is given as follows:

   \begin{itemize}
        \item[-] $((W,R,V), w)\models\smldiamond^{\psi}_{\chi}\varphi \textit{ iff there is an edge }(u,v)\in R \textit{ such that } ((W,R,V), u)\models\psi, ((W,R,V), v)\models\chi$ 
        
        $\textit{ and } ((W,R\backslash\{(u,v)\}, V),w)\models\varphi$
    \end{itemize}
    
    \noindent SML can be seen as a restricted version of GSML, where $\smldiamond^{\psi}_{\chi}\varphi$ is replaced by the formula $\smldiamond\varphi$ (with $\psi$ and $\chi$ replaced by $\top$) whose intuitive meaning is as follows: after some edge is deleted 
 from the model, the formula $\varphi$ still holds at the point of evaluation. The truth definition for $\smldiamond\varphi$ is given as follows: 
 
  \begin{itemize}
        \item[-] $((W,R,V), w)\models\smldiamond\varphi \textit{ iff there is an edge }(u,v)\in R \textit{ such that } ((W,R\backslash\{(u,v)\}, V),w)\models\varphi$
    \end{itemize}
   
\noindent         We note that for the truth definition of the modality "$\smldiamond$", we consider two different pointed models. We say that a formula $\smldiamond\varphi$ is true in a pointed model, if $\varphi$ is true in another pointed model. The latter model is not independent of the former one, and is related in a special way. Let us formalize this relation: Two pointed models, $(\mathcal{M}_1,w_1)=((W_1,R_1,V_1),w_1)$ and $(\mathcal{M}_2,w_2)=((W_2,R_2,V_2),w_2)$ are related by an "\textbf{r}" relation, or $((\mathcal{M}_1,w_1),(\mathcal{M}_2,w_2))\in \textbf{r}$ if the following holds: (i) $W_2 = W_1$, (ii) $R_2 = R_1\backslash\{(u,v)\}$ for some $(u,v)\in R_1$, (iii) $V_2 = V_1$, and (iv) $w_2 = w_1$. With this new relation, the semantics of $\smldiamond\varphi$ can be seen as follows:
  $(\mathcal{M},w)\models\smldiamond\varphi \textit{ iff there is some } (\mathcal{M}^\prime,w^\prime) \textit{ such that } (\mathcal{M},w)\textbf{r}(\mathcal{M}^\prime,w^\prime) \textit{ and } (\mathcal{M}^\prime,w^\prime)\models\varphi$.

       \medskip
        
        \noindent Let us now focus on the following question: When do two pointed models satisfy the same sabotage modal formulas? The relevant model-theoretic notion is that of sabotage bisimulation introduced in \cite{sab-logic}, given as follows:
          Let $\mathcal{M}_1=((W_1,R_1,V_1),w_1)$ and $\mathcal{M}_2=((W_2,R_2,V_2),w_2)$ be two pointed models. We say that 
                $(\mathcal{M}_1, w_1)$ and $(\mathcal{M}_2, w_2)$ are \emph{sabotage bisimilar}, denoted by $(\mathcal{M}_1,w_1)Z(\mathcal{M}_2,w_2)$, if the following conditions are satisfied:
                
                \begin{enumerate}
                    \item \textbf{Atom}: If $(\mathcal{M}_1,w_1)Z(\mathcal{M}_2,w_2)$, then $(\mathcal{M}_1,w_1)\models p$ iff $(\mathcal{M}_2,w_2)\models p$ for all atomic propositions $p$.
                    \item \textbf{Zig$_\mldiamond$}: If $(\mathcal{M}_1,w_1)Z(\mathcal{M}_2,w_2)$, and there exists $v_1\in W_1$ such that $w_1R_1v_1$, then there is a $v_2\in W_2$ such that $w_2R_2v_2$ and $(\mathcal{M}_1,v_1)Z(\mathcal{M}_2,v_2)$.
                   \item \textbf{Zag$_\mldiamond$}: Same as above in the converse direction.
                    \item \textbf{Zig$_\smldiamond$}: If $(\mathcal{M}_1,w_1)Z(\mathcal{M}_2,w_2)$, and there exists $\mathcal{M}_1^\prime$ such that $(\mathcal{M}_1,w_1){\bf r}(\mathcal{M}_1^\prime,w_1)$, then there is an $\mathcal{M}_2^\prime$ such that $(\mathcal{M}_2,w_2){\bf r}(\mathcal{M}_2^\prime,w_2)$ and $(\mathcal{M}_1^\prime,w_1)Z(\mathcal{M}_2^\prime,w_2)$.
                    \item  \textbf{Zag$_\smldiamond$}: Same as above in the converse direction.
                \end{enumerate}
                
                \begin{proposition}
	If two models $(\mathcal{M}_1,w_1)$, $(\mathcal{M}_2,w_2)$ are sabotage bisimilar, then they satisfy the same SML formulas.
            \end{proposition}

    \noindent Similarly, we say that two pointed models
                $(\mathcal{M}_1, w_1)$ and $(\mathcal{M}_2, w_2)$ are generalized sabotage bisimilar, denoted by $(\mathcal{M}_1,w_1)Z(\mathcal{M}_2,w_2)$, if conditions (1.), (2.), (3.) are as above and the following are satisfied:
                
                \begin{enumerate}
                    \item[4.] \textbf{Zig$_\smldiamond$}: If $(\mathcal{M}_1,w_1)Z(\mathcal{M}_2,w_2)$, and there exists $\mathcal{M}_1^\prime = (W_1,R_1\backslash\{(u_1,v_1)\},V_1)$ such that $(\mathcal{M}_1,w_1){\bf r}(\mathcal{M}_1^\prime,w_1)$, then there is an $\mathcal{M}_2^\prime= (W_2,R_2\backslash\{(u_2,v_2)\},V_2)$ such that $(\mathcal{M}_1,u_1)Z(\mathcal{M}_2,u_2)$, $(\mathcal{M}_1,v_1)Z(\mathcal{M}_2,v_2)$, $(\mathcal{M}_2,w_2){\bf r}$
                    
                    $(\mathcal{M}_2^\prime,w_2)$ and $(\mathcal{M}_1^\prime,w_1)Z(\mathcal{M}_2^\prime,w_2)$
                    \item[5.] \textbf{Zag$_\smldiamond$}: Same as above in the converse direction.
                    
                \end{enumerate}
                
                \begin{proposition}
	If two models $(\mathcal{M}_1,w_1)$ and $(\mathcal{M}_2,w_2)$ are generalized sabotage bisimilar, then they satisfy the same GSML formulas.
            \end{proposition}

\noindent We note that SML describes arbitrary link deletion, whereas GSML describes link deletion where the end-points of the links satisfy certain properties.

\subsection{Modal logics for stepwise removal}  

We now provide the language and semantics of MLSR as given in \cite{point-logic}. 
Given a countable set of propositional variables $\mathcal{P}$, the formulas of MLSR are given as follows:

      \begin{quote}
       $\varphi\coloneqq p\mid\neg\varphi\mid(\varphi\land\varphi)\mid\mldiamond\varphi\mid\langle-\varphi\rangle\varphi$
    \end{quote}
    
     \noindent  where $p\in\mathcal{P}$. The intuitive meaning of the formula $\langle-\psi\rangle\varphi$ is as follows: after some point that is distinct from the point of evaluation and where $\psi$ holds, is deleted 
 from the model, the formula $\varphi$ still holds at the point of evaluation. The models for MLSR are given by  standard relational models $\mathcal{M} = (W, R, V)$ for modal logics.
The truth definition 
of MLSR formulas in such pointed models given by $((W,R,V), w)$ are given as usual for the propositional, boolean and the modal formulas. For the removal modal operator, it is
given as follows:

    \begin{itemize}
        \item[-] $((W,R,V), w)\models\langle-\psi\rangle\varphi$ \textit{ iff there is a world $v\not= w$ such that}  $((W,R,V), v)\models\psi$ \textit{ and } $((W\setminus\{v\}, R^\prime, V^\prime),w)\models\varphi$
    \end{itemize}
    
    \noindent where, $R^\prime$ is a sub-relation of $R$ formed by removing all the edges incident with $v$, and for all $p\in\mathcal{P}$, $V^\prime(p) = V(p) \setminus \{v\}$.  Let point sabotage logic (PSL) denote a restricted version of MLSR, where the only difference is in the formula $\langle-\psi\rangle\varphi$ which is replaced by the formula $\langle-\rangle\varphi$ (with $\psi$ replaced by $\top$) whose intuitive meaning is as follows: after some point that is distinct from the point of evaluation, is deleted 
 from the model, the formula $\varphi$ still holds at the point of evaluation. The truth definition for $\langle-\rangle\varphi$ is given as follows: 
 
 \begin{itemize}
        \item[-] $((W,R,V), w)\models\langle-\rangle\varphi$ \textit{ iff there is a world $v\not= w$ such that} $((W\setminus\{v\}, R^\prime, V^\prime),w)\models\varphi$  
    \end{itemize}

\noindent        Here, $R^\prime$ and $V^\prime$ are defined as above.  Let us now focus on the following question: When do two pointed models satisfy the same point sabotage modal formulas? The relevant model-theoretic notion is that of point sabotage bisimulation, 
derived from a similar notion introduced in \cite{point-logic}, given as follows:
          First of all, if $\mathcal{M}=(W,R,V)$, then define $\mathcal{M}\setminus\{v\}$ as $((W\setminus\{v\}, R^\prime, V^\prime)$ given above. Let $\mathcal{M}_1=((W_1,R_1,V_1),w_1)$ and $\mathcal{M}_2=((W_2,R_2,V_2),w_2)$ be two pointed models. We say that 
                $(\mathcal{M}_1, w_1)$ and $(\mathcal{M}_2, w_2)$ are point sabotage bisimilar, denoted by $(\mathcal{M}_1,w_1)Z(\mathcal{M}_2,w_2)$, if conditions (1.), (2.), (3.) are as above and the following are satisfied:
                
                \begin{itemize}
                \item[4.] \textbf{Zig$_{\langle-\rangle}$}: If $(\mathcal{M}_1,w_1)Z(\mathcal{M}_2,w_2)$, and $u_1\in \mathcal{M}_1$ with $u_1\not=w_1$, then there is a $u_2\in \mathcal{M}_2$ such that $u_2\not= w_2$ and $(\mathcal{M}_1\setminus\{u_1\},w_1)Z(\mathcal{M}_2\setminus\{u_2\},w_2)$.
                \item[5.] \textbf{Zag$_{\langle-\rangle}$}: Same as above in the converse direction.
                    \end{itemize}

                \begin{proposition}
	If two models $(\mathcal{M}_1,w_1)$, $(\mathcal{M}_2,w_2)$ are point sabotage bisimilar, they satisfy the same PSL formulas.
            \end{proposition}

   \noindent We say that two pointed models
                $(\mathcal{M}_1, w_1)$ and $(\mathcal{M}_2, w_2)$ are MLSR bisimilar or generalized point sabotage bisimilar, denoted by $(\mathcal{M}_1,w_1)Z(\mathcal{M}_2,w_2)$, if conditions (1.), (2.), (3.) are as above and the following are satisfied:
                
                \begin{itemize}
                \item[4.] \textbf{Zig$_{\langle-\rangle}$}: If $(\mathcal{M}_1,w_1)Z(\mathcal{M}_2,w_2)$, and $u_1\in \mathcal{M}_1$ with $u_1\not=w_1$, then there is a $u_2\in \mathcal{M}_2$ such that $u_2\not= w_2$, $(\mathcal{M}_1,u_1)Z(\mathcal{M}_2,u_2)$ and $(\mathcal{M}_1\setminus\{u_1\},w_1)Z(\mathcal{M}_2\setminus\{u_2\},w_2)$.
                \item[5.] \textbf{Zag$_{\langle-\rangle}$}: Same as above in the converse direction.
                    \end{itemize}

                \begin{proposition}
	If two models $(\mathcal{M}_1,w_1)$ and $(\mathcal{M}_2,w_2)$ are generalized point sabotage bisimilar, then they satisfy the same MLSR formulas.
            \end{proposition}
            
            
            \noindent \noindent We note that PSL describes arbitrary point deletion, whereas MLSR describes point deletion satisfying certain properties. Before finishing this section, we should mention here that the only new concept that has been introduced here is that of generalized sabotage modal bisimulation. The rest are all taken up from the existing literature. With these four distinct notions of bisimulation, we are now all set to investigate the complexity of the following decision problem: Given two relational models, are they bisimilar?
            
 
  \section{Expressing bisimulations}  
   
      To get further insight into these distinct notions of model comparison or bisimulation, we now express them in their respective languages. For this study we assume finite models and finite propositional atoms which will be essential for our algorithmic studies of bisimulation. In what follows, we concentrate on sabotage bisimulation and point sabotage bisimulation. For discussion on their generalized versions see Appendix A.
      
   \subsection{Sabotage bisimulation (s-bisimulation)}
   
   Following \cite{old-paper} we now provide a characterizing formula corresponding to a finite pointed model which will be satisfied by any model bisimilar to it. Let us first 
define the formula $E(M)$ for any model $M$:
Take a set of new proposition letters (different from those in the initial language) $p_x$ for each world $x$ in $M$. Moreover, for each $x$, let $AT_x$ be the conjuction of all literals in the original language that hold at $M,x$. Then $E(M)$ is the conjuction of all statements $p_x\rightarrow AT_x\wedge env(M,x)$, where $env(M,x)$ is the formula $\bigwedge_{y\in \{y|R_{M} xy\}}\Diamond p_{y} \wedge \Box\bigvee_{y\in \{y|R_{M} xy\}} p_{y}$, where $R_M$ is the relation of model $M$. Suppose there are $n$ edges in $M$, that is, $i_1,...\ ,i_n$.  $M{(i_1,\ ...\ , i_k)}$ denote the new model after we delete edges $i_1,...\ ,i_k$. Let $P{(i_1,...,i_k)}$ denote $(i_1,\ ...\ ,i_k)\in R_M^{k}\And i_a\not=i_b,\ \text{if}\ a\not=b$. For $1\leq k\leq n$, let  $G_k(M)$ denote: 

$$ \bigwedge_{P_{(i_1,...,i_k)}} \blacklozenge^{k}E(M_{(i_1,\ ...\ , i_k)})\ \bigwedge \blacksquare^{k}\bigvee_{P_{(i_1,...,i_k)}} E(M_{(i_1,\ ...\ , i_k)})$$
Let $G_{n+1}(M)$ be $\neg \blacklozenge^{n+1}\top$, $G(M)$ be $E(M)\wedge\bigwedge_{1\leq i\leq n+1}G_i(M)$. Intuitively, $P_{(i_1,...,i_k)}$ means the collection of link sequences which consists of $k$ links different from each other. If $G(M)$ is valid on some model, the new  model after some link-cutting sequence maintains similarity with the corresponding model  after some link-cutting sequence. Especially, the case where the link cutting sequence is empty, is the case of basic modal logic. The following theorem provides a connection between the s-bisimulation and the formula described above. Let the symbol $\underline{\leftrightarrow}_s$ indicate the existence of an s-bisimulation between two pointed models.
\begin{theorem}
	For any two pointed models $(M=(W_1,R_1,V_1),s)$, $(N=(W_2,R_2,V_2),t)$, the following are equivalent:
	\begin{itemize}
	\item[$(a)$] $(M,s)\underline{\leftrightarrow}_s (N,t)$
	\item[$(b)$] $(N,t)$ can be expanded to a model $(N^\prime,t)$ for $G(M)$ such that $p_s$ holds at $t$.
\end{itemize}
\end{theorem}
\begin{proof}
	$[(a)\Longrightarrow(b)]$  Define $V_{2}^{\prime}(p)= V_2 (p)$ for any proposition letter in the initial language,   $V_{2}^{\prime}(p_x)=\{u|(M,x)\underline{\leftrightarrow}_s$
	
\noindent	$ (N,u) \}$ for $x\in W_1$. We prove that under this valuation, 
	$N^\prime\vDash G(M)\And N^\prime, t\vDash p_{s}$. Since  $(M,s)\underline{\leftrightarrow}_s (N,t)$, then $t\in V_{2}^{\prime}(p_s)$, which means $p_s$ holds at world $t$. We have to prove $N^\prime\vDash G(M)$, where $G(M)$ is $E(M)\wedge\bigwedge_{1\leq i\leq n+1}G_i(M)$. Firstly, we prove that  $N^\prime\vDash E(M)$ by contradiction.
		
     $E(M)$ is the conjuction of all statements $p_x\rightarrow AT_x\wedge env(M,x)$, where $env(M,x)$ is the formula $\bigwedge_{y\in \{y|R_1 xy\}}\Diamond p_{y} \wedge \Box\bigvee_{y\in \{y|R_1 xy\}} p_{y}$.
	Suppose $N^\prime\nvDash E(M)$, then there exists $e$ in $N^\prime$ such that $N^\prime, e \nvDash p_y\rightarrow AT_y \wedge env(M,y)$ for some $y\in W_1$, then $N^\prime, e \vDash p_y$ and $N^\prime, e \nvDash AT_y \wedge env(M,y)$. Since $N^\prime, e \vDash p_y$, then $(M,y)\underline{\leftrightarrow}_s (N,e)$, then $(M,y)\vDash p$ iff $N,e\vDash p$ for any proposition letter $p$ in the initial language.
	\begin{itemize}
		\item [$-$] If $N^\prime, e \nvDash AT_y$, then there exists $p$ in the initial language such that $(M,y)\vDash p$ iff $N^\prime,e\nvDash p$ iff $N,e\nvDash p$, contradiction.
		\item [$-$] If $N^\prime, e \nvDash  env(M,y)$, then $N^\prime, e \nvDash \bigwedge_{z\in \{y|R_{1} yz\}}\Diamond p_{z} \wedge \Box\bigvee_{z\in \{y|R_{1} yz\}} p_{z}$.
		\begin{itemize}
			\item[$-$] If $N^\prime, e \nvDash \bigwedge_{z\in \{y|R_{1} yz\}}\Diamond p_{z} $, then  $N^\prime, e \nvDash \Diamond p_z$ for some $z$ with $R_{1}yz$, since $(M,y)\underline{\leftrightarrow}_s (N,e)$, then there exists $z^\prime$ with $R_2e{z}^\prime$ and $(M,z)\underline{\leftrightarrow}_s (N,z^\prime)$, then $z^\prime\in V_{2^{\prime}}(p_z)$, thus $N,z^\prime\vDash p_z$, then we have $N,e\vDash\Diamond p_z$, contradiction.
			\item[$-$] If $N^\prime, e \nvDash \Box\bigvee_{z\in \{z|R_{1} yz\}} p_{z}$, then there exists world $e^\prime$ with $R_{2}ee^\prime$, such that $N^\prime, e^\prime \vDash \bigwedge_{z\in \{y|R_{1} yz\}}\neg p_z$. since $(M,y)\underline{\leftrightarrow}_s (N,e)$, then there exists $y^\prime$ with $R_{1}yy^\prime$ such that $(M,y^\prime)\underline{\leftrightarrow}_s (N,e^\prime)$, then $p_{y^\prime}$ holds at world $e^\prime$, contradiction.
		\end{itemize}
	\end{itemize}
Then we prove $N^\prime\vDash\bigwedge_{1\leq i\leq n+1}G_i(M)$ directly. For any  link-cutting sequence $(d_1,...,d_l)$ satisfying $P_{(d_1,...,d_l)}$ from  $M$, because of the $Zig_{\blacklozenge}$ item of s-bisimulation, there exists a link-cutting sequence $(j_1,...,j_l)$ satisfying $P_{(j_1,...,j_l)}$ from  $N$, such that  $(M_{(d_1,...,d_l)},s)\underline{\leftrightarrow}_s (N_{(j_1,...,j_l)},t)$. Similarly with the first part, we have that $(N_{(j_1,...,j_l)},t)$ can be expanded to a model $({N_{(j_1,...,j_l)}}^\prime,t)$ for $E(M_{(d_1,...,d_l)})$, since ${N_{(j_1,...,j_l)}}^\prime= N_{(j_1,...,j_l)}^\prime$, then $ N^\prime\vDash \blacklozenge^{l}E(M_{(d_1,...,d_l)})$, thus $ N^\prime\vDash \bigwedge_{P_{(i_1,...,i_l)}} \blacklozenge^{l}E(M_{(i_1,\ ...\ , i_l)})$. Since for any link-cutting sequence $(j_1,...,j_l)$ satisfying $P_{(j_1,...,j_l)}$, we have $ N_{(j_1,...,j_l)}^\prime\vDash \bigvee_{P_{(i_1,...,i_l)}} E(M_{(i_1,\ ...\ , i_l)})$, then $ N^\prime\vDash \blacksquare^{l}\bigvee_{P_{(i_1,...,i_l)}} E(M_{(i_1,\ ...\ , i_l)})$. And since model $N$ has the same number of edges as model $M$ (cf. Lemma 20, Appendix B), $N^\prime\vDash \neg \blacklozenge^{n+1}\top$, and hence $N^\prime\vDash G(M)$.

\smallskip

$[(b)\Longrightarrow(a)]$ To prove this direction it is sufficient to prove Lemmas $6$ and $7$ given below. The atomic case is taken care of by $G(M)$. Lemma $6$ is used to show the $\mathit{Zig}$ condition for  $\Diamond$ and $\blacklozenge$, while lemma $7$ is used for the corresponding $\mathit{Zag}$ conditions. 
\end{proof} 

	\begin{lemma}
		For any link-cutting sequence $(d_1,...,d_l)$ satisfying $P_{(d_1,...,d_l)}$ from  $M$, where $0\leq l\leq n$, if there is a  path $s_0\rightarrow s_1\rightarrow\ ...\ \rightarrow s_k$ in $M_{(d_1,...,d_l)}$, then there exists a link-cutting sequence $(j_1,...,j_l)$ satisfying $P_{(j_1,...,j_l)}$ from  $N$ and a path  $t_0\rightarrow t_1\rightarrow\ ...\ \rightarrow t_k$ in $N_{(j_1,...,j_l)}$ such that $s_i\in V_{1}(p)$ iff $t_i\in V_{2}(p)$ for each proposition letter $p$ in the initial language, where $s_0$ is reachable from $s$ in $M$, $t_0$ is reachable from $t$ in $N$, $i\in [0,k]$.
	\end{lemma}
	\begin{proof}
		For any path $s_0\rightarrow s_1\rightarrow\ ...\ \rightarrow s_k$ in $M_{(d_1,...,d_l)}$, since $ N^\prime\vDash \bigwedge_{P_{(i_1,...,i_l)}} \blacklozenge^{l}E(M_{(i_1,\ ...\ , i_l)})$, then $ N^\prime\vDash \blacklozenge^{l}E(M_{(d_1,\ ...\ , d_l)})$, thus  $ N^\prime_{(j_1,...,j_l)}\vDash E(M_{(d_1,\ ...\ , d_l)})$ for some link-cutting sequence $(j_1,...,j_l)$ from  $N^\prime$. We take $s_0$ satisfying $N^\prime,t_0\vDash p_{s_0}$, since $N^\prime_{(j_1,...,j_l)}, t_0\vDash p_{s_0}\rightarrow AT_{s_0} \wedge env(M_{(d_1,...,d_l)},s_0)$, then $N^\prime,t_0\vDash AT_{s_0}$. Suppose that we have selected $t_i (i<k)$ satisfying $N_{(j_1,...,j_l)}^\prime,t_i\vDash  p_{s_i}$, we prove that there exists $t_{i+1}$ such that $N^\prime_{(j_1,...,j_l)}, t_{i+1}\vDash p_{s_{i+1}}$ and $R_{2}t_{i}t_{i+1}$. Since $N^\prime_{(j_1,...,j_l)}, t_i\vDash p_{s_i}\rightarrow AT_{s_i} \wedge env(M_{(d_1,...,d_l)},s_i)$, then $N^\prime_{(j_1,...,j_l)},t_i\vDash AT_{s_i}$, and $N^\prime_{(j_1,...,j_l)},t_i\vDash env(M_{(d_1,...,d_l)},s_i)$. Thus we have $N^\prime_{(j_1,...,j_l)},t_i\vDash \Diamond p_{s_{i+1}}$, then  there exists $e$ with $R_{2} t_{i}e$ such that $N^\prime_{(j_1,...,j_l)},e\vDash p_{s_{i+1}}$, since 	$N^\prime_{(j_1,...,j_l)}, e\vDash p_{s_{i+1}}\rightarrow AT_{s_{i+1}} \wedge env(M_{(d_1,...,d_l)},s_{i+1})$, then 	$N^\prime_{(j_1,...,j_l)}, e\vDash AT_{s_{i+1}}$. Let $t_{i+1}$ be $e$, then we finish our lemma.	
	\end{proof}
	
	\begin{lemma}
		For any link-cutting sequence $(j_1,...,j_l)$ satisfying $P_{(j_1,...,j_l)}$ from  $N$, where $0\leq l\leq n$, if there is a path $t_0\rightarrow t_1\rightarrow\ ...\ \rightarrow t_k$ in $N_{(j_1,...,j_l)}$ , then there exists a link-cutting sequence $(d_1,...,d_l)$  satisfying $P_{(d_1,...,d_l)}$ from $M$ and a path $s_0\rightarrow s_1\rightarrow\ ...\ \rightarrow s_k$ in $M_{(d_1,...,d_l)}$  such that $s_i\in V_{1}(p)$ iff $t_i\in V_{2}(p)$ for each proposition letter $p$ in the initial language, where $t_0$ is reachable from $t$ in $N$, $s_0$ is reachable from $s$ in $M$, $i\in [0,k]$.
	\end{lemma}
	
	\begin{proof}	
		Since $ N^\prime\vDash \blacksquare^{l}\bigvee_{P_{(i_1,...,i_l)}} E(M_{(i_1,\ ...\ , i_l)})$, then for any link-cutting sequence $(j_1,...,j_l)$ from  $N$,  $N_{(j_1,...,j_l)}^\prime \vDash E(M_{(d_1,...,d_l)})$ for some link-cutting sequence $(d_1,...,d_l)$ from  $M$. For any path $t_0\rightarrow t_1\rightarrow\ ...\ \rightarrow t_k$ in $N_{(j_1,...,j_l)}$, we select $s_0$ satisfing $N_{(j_1,...,j_l)}^\prime,t_0\vDash p_{s_{0}}$, since 
		$N^\prime_{(j_1,...,j_l)}, t_0\vDash p_{s_0}\rightarrow AT_{s_0} \wedge env(M_{(d_1,...,d_l)},s_0)$, then $N^\prime_{(j_1,...,j_l)},t_0\vDash AT_{s_0}$.  Suppose that we have selected $s_i (i<k)$ satisfying $N_{(j_1,...,j_l)}^\prime,t_i\vDash  p_{s_i}$, we prove that there exists $s_{i+1}$ such that $N^\prime, t_{i+1}\vDash p_{s_{i+1}}$ and $R_{1}s_{i}s_{i+1}$. Since $N^\prime_{(j_1,...,j_l)}, t_i\vDash p_{s_i}\rightarrow AT_{s_i} \wedge env(M_{(d_1,...,d_l)},s_i)$, then $N^\prime_{(j_1,...,j_l)},t_i\vDash AT_{s_i}$, and $N^\prime_{(j_1,...,j_l)},t_i\vDash env(M_{(d_1,...,d_l)},s_i)$. If $env(M_{(d_1,...,d_l)},s_i)$ is $\top\wedge\Box\bot$(when $s_i$ has no successor in $M_{(d_1,...,d_l)}$),  then $N^\prime_{(j_1,...,j_l)},t_i\vDash\Box\bot$, it's impossible since $t_{i+1}$ is a successor of world $t_i$ in $N^\prime_{(j_1,...,j_l)}$, then there exists $z$ with $R_{1}s_{i}z$ such that $N^\prime_{(j_1,...,j_l)}, t_{i+1}\vDash p_{z}$, since 
		$N^\prime_{(j_1,...,j_l)}, t_{i+1}\vDash p_{z}\rightarrow AT_{z} \wedge env(M_{(d_1,...,d_l)},z)$, then $N^\prime_{(j_1,...,j_l)}, t_{i+1}\vDash  AT_{z}$. Let $s_{i+1}$ be  $z$, then we finish our proof.
	\end{proof}

	\noindent If the cutting sequence is empty, then the above lemmas reduce to the case of basic modal logic as follows:
	
\begin{proposition}
	The following conditions are equivalent:\\
	$(a)$ $(M,s)$ is bisimilar to $(N,t)$.\\
	$(b)$ For any path $s_0\rightarrow s_1\rightarrow\ ...\ \rightarrow s_k$ in $M$ there is a path $t_0\rightarrow t_1\rightarrow\ ...\ \rightarrow t_k$ in $N$, and for any path  $t_0\rightarrow t_1\rightarrow\ ...\ \rightarrow t_k$ in $N$, there is a path $s_0\rightarrow s_1\rightarrow\ ...\ \rightarrow s_k$ in $M$, such that $s_i\in V_{1}(p)$ iff $t_i\in V_{2}(p)$ for each proposition letter $p$.  Here,  $s_0$ is reachable from $s$ in M, and $t_0$ is reachable from $t$ in $N$.
\end{proposition} 


 \subsection{Point sabotage bisimulation (d-bisimulation)}
 
 Now we move on to prove a similar result for point sabotage bisimulation. Suppose that there are $n$ worlds in a model $M=(W,R,V)$, let $M_{[e_1,\ ...\ , e_k]}$ denote the new model after we delete the worlds $e_1,\ ...\ , e_k$. Let $Q_{(e_1,...,e_k)}$ denote $(e_1,\ ...\ ,e_k)\in W^{k}\And e_i\not=e_j,\ \text{if}\ i\not=j$. When $1\leq k <n$, let  $H_k(M)$ denote 

$$ \bigwedge_{Q_{(e_1,...,e_k)}} {\left\langle -\right\rangle }^{k}E(M_{[e_1,\ ...\ , e_k]})\ \bigwedge {\left[ -\right] }^{k}\bigvee_{Q_{(e_1,...,e_k)}} E(M_{[e_1,\ ...\ , e_k]})$$
Let $H_{n}(M)$ be $\neg{\left\langle -\right\rangle}^{n}\top$, $H(M)$ be $E(M)\wedge\bigwedge_{1\leq i\leq n}H_i(M)$. Intuitively, $Q_{(e_1,...,e_k)}$ means the class of point sequences which consist of $k$ points different from each other. If $H(M)$ is valid on some model, which means the new  model after some point-deleting sequence keeps similar to the corresponding model  after some point-deleting sequence. Especially, the case where the point-deleting sequence is empty, is the case of basic modal logic. The following theorem provides a connection between d-bisimulation and the formula described above. Let the symbol $\underline{\leftrightarrow}_d$ indicate the existence of a d-bisimulation between two pointed models.

\begin{theorem}
	For any two pointed models $(M,s)=((W_1,R_1,V_1),s)$, and $(N,t)= ((W_2,R_2,V_2),t)$, we have that the following are equivalent:
	\begin{itemize}
	\item[$(a)$] $(M,s)\underline{\leftrightarrow}_d (N,t)$
	\item[$(b)$] $(N,t)$ can be expanded to a model $(N^\prime,t)$ for $H(M)$ such that $p_s$ holds at world $t$.
	\end{itemize} 
\end{theorem}

\noindent The proof of this theorem is very similar to that of the result on s-bisimulation, and is given in details in Appendix A.
 A natural question that might arise here is the following: Why are we trying to prove these logical characterization results for bisimulations? These results which relate the existence of a bisimulation between two models to a certain model checking problem do shed some light on the complexity of these problems (see below). Additionally, they provide us with a structural representation of the whole (finite) model in terms of a formula in the respective logic which captures the essence of bisimulation as a notion of invariance for the corresponding modal logic.
 
 \smallskip
 
 \noindent To continue with our discussion on the complexity problem, let us now ask another general question: How hard is it to show that two finite models are bisimilar? The theorems 5 and 9 that we have shown above and the fact that the model checking problems for SML and MLSR are PSPACE-complete \cite{mfcs2003,point-logic}  suggest us an upper bound for the model comparison/bisimulation problems of the logics SML (link deletion logic) and PSL (point deletion logic). In the next section, we tackle these problems algorithmically and provide the complexity results accordingly. We note here that the results similar to theorems 5 and 9 that are proved in Appendix A for the logics GSML and MLSR, together with the complexity results mentioned above suggest an upper bound for the bisimulation problems of those logics as well.

\section{An algorithmic study}

Let us now provide algorithms to check whether two pointed models are bisimilar - we have four distinct notions of bisimilarity based on different logics of link and point deletion. Natural questions would be as follows: How are these algorithms connect to each other? Can one be reduced to the other? Before trying to answer such questions we provide algorithms to check bisimulation between pointed models for each of these logics. We note here that the algorithm corresponding to sabotage bisimulation is presented here, whereas the algorithms corresponding to the other three notions of bisimulation, which are similar in nature, are presented in Appendix B.

\subsection{Sabotage bisimulation (s-bisimulation)}

In what follows, we provide an algorithm (Algorithm 1 in Page 8) for checking sabotage bisimulation, prove its correctness and check the complexity. We define a function \textsf{s-Bisimilar} that takes as input two pointed relational models, $(\mathcal{M}_1, w_1)$, $(\mathcal{M}_2, w_2)$ and a list $L\subseteq W_1\times W_2$, where $\mathcal{M}_1=(W_1,R_1,V_1)$ and $\mathcal{M}_2=(W_2,R_2,V_2)$, and outputs "Yes" if the two models are s-bisimlar and the function is called with $L=\emptyset$, and "No" when the given models are not s-bisimilar and the function is called with $L=\emptyset$.  A  proof of correctness for Algorithm 1 is provided in Appendix B (it could not be included due to space restrictions). Below, we show that this algorithm is in PSPACE.

\bigskip

\begin{algorithm}[tbp]\footnotesize
            \LinesNumbered
                \KwIn{ $((W_1,R_1,V_1),w_1), ((W_2,R_2,V_2),w_2)$ } 
                \textbf{Initialize}:L=$\emptyset$\\
                \SetKwFunction{FMain}{s-Bisimilar}
                \SetKwProg{Fn}{Function}{:}{}
                \Fn{\FMain{$((W_1,R_1,V_1),w_1), ((W_2,R_2,V_2),w_2)$, L}}{
                    {\If{$|R_1|\neq |R_2|$}
                    {\Return NO;}
                    }

                    {
                        \ForAll{atomic propositions p}
                        {
                            \If{$(((w_1\in V_1(p)) AND (w_2\not\in V_2(p)))$ OR $((w_1\not\in V_1(p)) AND (w_2\in V_2(p))))$}
                                {
                                    \Return NO;
                                }
                        }
                        \ForAll{$e_1\in R_1$}
                            {
                                Found=0;\\
                                \ForAll{$e_2\in R_2$}
                                    {
                                        \If {\textsf{s-Bisimilar}$(((W_1,R_1\backslash\{e_1\},V_1),w_1)$,$((W_2,R_2\backslash\{e_2\},V_2),w_2), \emptyset)$==YES}
                                            {
                                                Increment Found;\\
                                                break;
                                            }
                                    }
                                \If{Found =0}
                                    {
                                        \Return No;
                                    }
                            }
                        
                        \ForAll{$e_2\in R_2$}
                            {
                                Found=0;\\
                                \ForAll{$e_1\in R_1$}
                                    {
                                        \If {\textsf{s-Bisimilar}$(((W_1,R_1\backslash\{e_1\},V_1),w_1),((W_2,R_2\backslash\{e_2\},V_2),w_2), \emptyset)$==YES 
                                        }
                                            {
                                                Increment Found;\\
                                                break;
                                            }
                                    }
                                \If{Found =0}
                                    {
                                        \Return No;
                                    }}

                            \If {$(w_1,w_2)\not\in L$}
                            {
                            \ForAll{$u_1\in W_1$}
                            {
                                Found=0;\\
                                \ForAll{$u_2\in W_2$}
                                    {
                                        \If {($(w_1R_1u_1)$ AND $(w_2R_2u_2))$}
                                        {
                                            \If{$((u_1,u_2)\not\in L)$}
                                            {
                                                \If {\textsf{s-Bisimilar}$(((W_1,R_1,V_1),u_1),((W_2,R_2,V_2),u_2),L\cup\{(w_1,w_2)\})$==YES}
                                                {
                                                    Increment Found;
                                                }
                                            }
                                            \Else{Increment Found}
                                        }
                                            
                                    }
                                \If{(Found=0) AND $(w_1R_1u_1)$}
                                    {
                                        \Return No;
                                    }
                            }
                            
                        \ForAll{$u_2\in W_2$}
                            {
                                Found=0;\\
                                \ForAll{$u_1\in W_1$}
                                    {
                                        \If {($(w_1R_1u_1)$ AND $(w_2R_2u_2))$}
                                        {
                                            \If{$((u_1,u_2)\not\in L)$}
                                            {
                                                \If {\textsf{s-Bisimilar}$(((W_1,R_1,V_1),u_1),((W_2,R_2,V_2),u_2),L\cup\{(w_1,w_2)\})$==YES}
                                                {
                                                    Increment Found;
                                                }
                                            }
                                            \Else{Increment Found}
                                        }
                                            
                                    }
                                \If{(Found=0) AND $(w_2R_2u_2)$}
                                    {
                                        \Return No;
                                    }
                            }
                        }
                        
                        \Return Yes;
                    }
                }
                \caption{Algorithm to check whether two models are s-bisimilar}
            \end{algorithm}

        \begin{theorem}
               Function \textsf{s-Bisimilar} terminates and is in PSPACE.
        \end{theorem}

        \begin{proof}
            We will form a recursion tree to see whether the function \textsf{s-Bisimilar} terminates and analyze the space complexity of the function.
            
                    \tikzset{every picture/.style={line width=0.75pt}} 

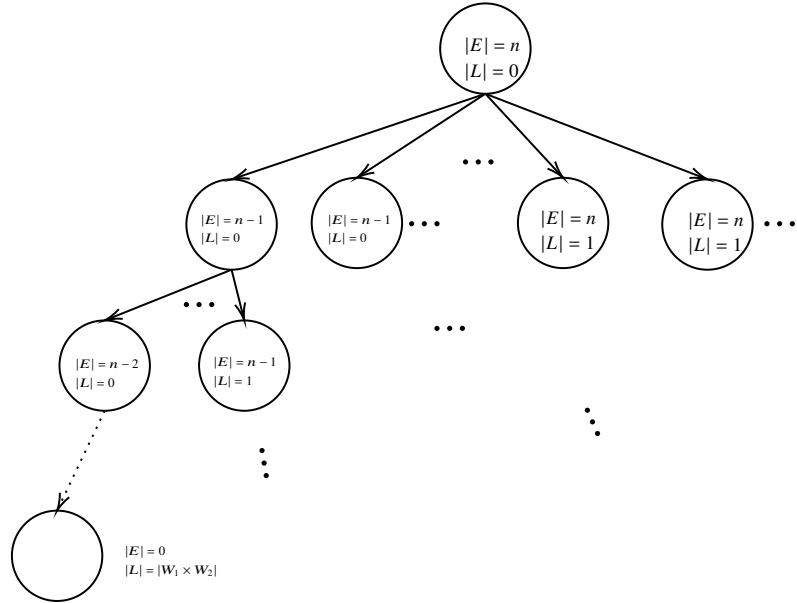
\begin{center}
\begin{tikzpicture}[x=0.75pt,y=0.75pt,yscale=-1,xscale=1,scale=0.8, every node/.style={scale=0.8}]]

\draw   (281,39.5) .. controls (281,23.76) and (293.76,11) .. (309.5,11) .. controls (325.24,11) and (338,23.76) .. (338,39.5) .. controls (338,55.24) and (325.24,68) .. (309.5,68) .. controls (293.76,68) and (281,55.24) .. (281,39.5) -- cycle ;
\draw   (121,150.5) .. controls (121,134.76) and (133.76,122) .. (149.5,122) .. controls (165.24,122) and (178,134.76) .. (178,150.5) .. controls (178,166.24) and (165.24,179) .. (149.5,179) .. controls (133.76,179) and (121,166.24) .. (121,150.5) -- cycle ;
\draw   (200,149.5) .. controls (200,133.76) and (212.76,121) .. (228.5,121) .. controls (244.24,121) and (257,133.76) .. (257,149.5) .. controls (257,165.24) and (244.24,178) .. (228.5,178) .. controls (212.76,178) and (200,165.24) .. (200,149.5) -- cycle ;
\draw   (330,149.5) .. controls (330,133.76) and (342.76,121) .. (358.5,121) .. controls (374.24,121) and (387,133.76) .. (387,149.5) .. controls (387,165.24) and (374.24,178) .. (358.5,178) .. controls (342.76,178) and (330,165.24) .. (330,149.5) -- cycle ;
\draw   (421,150.5) .. controls (421,134.76) and (433.76,122) .. (449.5,122) .. controls (465.24,122) and (478,134.76) .. (478,150.5) .. controls (478,166.24) and (465.24,179) .. (449.5,179) .. controls (433.76,179) and (421,166.24) .. (421,150.5) -- cycle ;
\draw   (41,239.5) .. controls (41,223.76) and (53.76,211) .. (69.5,211) .. controls (85.24,211) and (98,223.76) .. (98,239.5) .. controls (98,255.24) and (85.24,268) .. (69.5,268) .. controls (53.76,268) and (41,255.24) .. (41,239.5) -- cycle ;
\draw   (129,239.5) .. controls (129,223.76) and (141.76,211) .. (157.5,211) .. controls (173.24,211) and (186,223.76) .. (186,239.5) .. controls (186,255.24) and (173.24,268) .. (157.5,268) .. controls (141.76,268) and (129,255.24) .. (129,239.5) -- cycle ;
\draw   (11,359.5) .. controls (11,343.76) and (23.76,331) .. (39.5,331) .. controls (55.24,331) and (68,343.76) .. (68,359.5) .. controls (68,375.24) and (55.24,388) .. (39.5,388) .. controls (23.76,388) and (11,375.24) .. (11,359.5) -- cycle ;
\draw    (309.5,68) -- (151.39,121.36) ;
\draw [shift={(149.5,122)}, rotate = 341.35] [color={rgb, 255:red, 0; green, 0; blue, 0 }  ][line width=0.75]    (10.93,-3.29) .. controls (6.95,-1.4) and (3.31,-0.3) .. (0,0) .. controls (3.31,0.3) and (6.95,1.4) .. (10.93,3.29)   ;
\draw    (309.5,68) -- (230.17,119.9) ;
\draw [shift={(228.5,121)}, rotate = 326.8] [color={rgb, 255:red, 0; green, 0; blue, 0 }  ][line width=0.75]    (10.93,-3.29) .. controls (6.95,-1.4) and (3.31,-0.3) .. (0,0) .. controls (3.31,0.3) and (6.95,1.4) .. (10.93,3.29)   ;
\draw    (309.5,68) -- (357.14,119.53) ;
\draw [shift={(358.5,121)}, rotate = 227.25] [color={rgb, 255:red, 0; green, 0; blue, 0 }  ][line width=0.75]    (10.93,-3.29) .. controls (6.95,-1.4) and (3.31,-0.3) .. (0,0) .. controls (3.31,0.3) and (6.95,1.4) .. (10.93,3.29)   ;
\draw    (309.5,68) -- (447.63,121.28) ;
\draw [shift={(449.5,122)}, rotate = 201.09] [color={rgb, 255:red, 0; green, 0; blue, 0 }  ][line width=0.75]    (10.93,-3.29) .. controls (6.95,-1.4) and (3.31,-0.3) .. (0,0) .. controls (3.31,0.3) and (6.95,1.4) .. (10.93,3.29)   ;
\draw    (149.5,179) -- (71.36,210.26) ;
\draw [shift={(69.5,211)}, rotate = 338.2] [color={rgb, 255:red, 0; green, 0; blue, 0 }  ][line width=0.75]    (10.93,-3.29) .. controls (6.95,-1.4) and (3.31,-0.3) .. (0,0) .. controls (3.31,0.3) and (6.95,1.4) .. (10.93,3.29)   ;
\draw    (149.5,179) -- (157.01,209.06) ;
\draw [shift={(157.5,211)}, rotate = 255.95999999999998] [color={rgb, 255:red, 0; green, 0; blue, 0 }  ][line width=0.75]    (10.93,-3.29) .. controls (6.95,-1.4) and (3.31,-0.3) .. (0,0) .. controls (3.31,0.3) and (6.95,1.4) .. (10.93,3.29)   ;
\draw  [dash pattern={on 0.84pt off 2.51pt}]  (69.5,268) -- (40.36,329.19) ;
\draw [shift={(39.5,331)}, rotate = 295.46] [color={rgb, 255:red, 0; green, 0; blue, 0 }  ][line width=0.75]    (10.93,-3.29) .. controls (6.95,-1.4) and (3.31,-0.3) .. (0,0) .. controls (3.31,0.3) and (6.95,1.4) .. (10.93,3.29)   ;
\draw  [fill={rgb, 255:red, 0; green, 0; blue, 0 }  ,fill opacity=1 ] (262,150) .. controls (262,149.45) and (262.45,149) .. (263,149) .. controls (263.55,149) and (264,149.45) .. (264,150) .. controls (264,150.55) and (263.55,151) .. (263,151) .. controls (262.45,151) and (262,150.55) .. (262,150) -- cycle ;
\draw  [fill={rgb, 255:red, 0; green, 0; blue, 0 }  ,fill opacity=1 ] (270,150) .. controls (270,149.45) and (270.45,149) .. (271,149) .. controls (271.55,149) and (272,149.45) .. (272,150) .. controls (272,150.55) and (271.55,151) .. (271,151) .. controls (270.45,151) and (270,150.55) .. (270,150) -- cycle ;
\draw  [fill={rgb, 255:red, 0; green, 0; blue, 0 }  ,fill opacity=1 ] (278,150) .. controls (278,149.45) and (278.45,149) .. (279,149) .. controls (279.55,149) and (280,149.45) .. (280,150) .. controls (280,150.55) and (279.55,151) .. (279,151) .. controls (278.45,151) and (278,150.55) .. (278,150) -- cycle ;
\draw  [fill={rgb, 255:red, 0; green, 0; blue, 0 }  ,fill opacity=1 ] (486,150) .. controls (486,149.45) and (486.45,149) .. (487,149) .. controls (487.55,149) and (488,149.45) .. (488,150) .. controls (488,150.55) and (487.55,151) .. (487,151) .. controls (486.45,151) and (486,150.55) .. (486,150) -- cycle ;
\draw  [fill={rgb, 255:red, 0; green, 0; blue, 0 }  ,fill opacity=1 ] (494,150) .. controls (494,149.45) and (494.45,149) .. (495,149) .. controls (495.55,149) and (496,149.45) .. (496,150) .. controls (496,150.55) and (495.55,151) .. (495,151) .. controls (494.45,151) and (494,150.55) .. (494,150) -- cycle ;
\draw  [fill={rgb, 255:red, 0; green, 0; blue, 0 }  ,fill opacity=1 ] (502,150) .. controls (502,149.45) and (502.45,149) .. (503,149) .. controls (503.55,149) and (504,149.45) .. (504,150) .. controls (504,150.55) and (503.55,151) .. (503,151) .. controls (502.45,151) and (502,150.55) .. (502,150) -- cycle ;
\draw  [fill={rgb, 255:red, 0; green, 0; blue, 0 }  ,fill opacity=1 ] (296,111) .. controls (296,110.45) and (296.45,110) .. (297,110) .. controls (297.55,110) and (298,110.45) .. (298,111) .. controls (298,111.55) and (297.55,112) .. (297,112) .. controls (296.45,112) and (296,111.55) .. (296,111) -- cycle ;
\draw  [fill={rgb, 255:red, 0; green, 0; blue, 0 }  ,fill opacity=1 ] (304,111) .. controls (304,110.45) and (304.45,110) .. (305,110) .. controls (305.55,110) and (306,110.45) .. (306,111) .. controls (306,111.55) and (305.55,112) .. (305,112) .. controls (304.45,112) and (304,111.55) .. (304,111) -- cycle ;
\draw  [fill={rgb, 255:red, 0; green, 0; blue, 0 }  ,fill opacity=1 ] (312,111) .. controls (312,110.45) and (312.45,110) .. (313,110) .. controls (313.55,110) and (314,110.45) .. (314,111) .. controls (314,111.55) and (313.55,112) .. (313,112) .. controls (312.45,112) and (312,111.55) .. (312,111) -- cycle ;
\draw  [fill={rgb, 255:red, 0; green, 0; blue, 0 }  ,fill opacity=1 ] (120,201) .. controls (120,200.45) and (120.45,200) .. (121,200) .. controls (121.55,200) and (122,200.45) .. (122,201) .. controls (122,201.55) and (121.55,202) .. (121,202) .. controls (120.45,202) and (120,201.55) .. (120,201) -- cycle ;
\draw  [fill={rgb, 255:red, 0; green, 0; blue, 0 }  ,fill opacity=1 ] (128,201) .. controls (128,200.45) and (128.45,200) .. (129,200) .. controls (129.55,200) and (130,200.45) .. (130,201) .. controls (130,201.55) and (129.55,202) .. (129,202) .. controls (128.45,202) and (128,201.55) .. (128,201) -- cycle ;
\draw  [fill={rgb, 255:red, 0; green, 0; blue, 0 }  ,fill opacity=1 ] (136,201) .. controls (136,200.45) and (136.45,200) .. (137,200) .. controls (137.55,200) and (138,200.45) .. (138,201) .. controls (138,201.55) and (137.55,202) .. (137,202) .. controls (136.45,202) and (136,201.55) .. (136,201) -- cycle ;
\draw  [fill={rgb, 255:red, 0; green, 0; blue, 0 }  ,fill opacity=1 ] (278,216) .. controls (278,215.45) and (278.45,215) .. (279,215) .. controls (279.55,215) and (280,215.45) .. (280,216) .. controls (280,216.55) and (279.55,217) .. (279,217) .. controls (278.45,217) and (278,216.55) .. (278,216) -- cycle ;
\draw  [fill={rgb, 255:red, 0; green, 0; blue, 0 }  ,fill opacity=1 ] (286,216) .. controls (286,215.45) and (286.45,215) .. (287,215) .. controls (287.55,215) and (288,215.45) .. (288,216) .. controls (288,216.55) and (287.55,217) .. (287,217) .. controls (286.45,217) and (286,216.55) .. (286,216) -- cycle ;
\draw  [fill={rgb, 255:red, 0; green, 0; blue, 0 }  ,fill opacity=1 ] (294,216) .. controls (294,215.45) and (294.45,215) .. (295,215) .. controls (295.55,215) and (296,215.45) .. (296,216) .. controls (296,216.55) and (295.55,217) .. (295,217) .. controls (294.45,217) and (294,216.55) .. (294,216) -- cycle ;
\draw  [fill={rgb, 255:red, 0; green, 0; blue, 0 }  ,fill opacity=1 ] (168.51,292.12) .. controls (169.05,292.03) and (169.57,292.4) .. (169.66,292.95) .. controls (169.75,293.49) and (169.38,294.01) .. (168.84,294.1) .. controls (168.29,294.19) and (167.78,293.82) .. (167.69,293.28) .. controls (167.59,292.73) and (167.96,292.22) .. (168.51,292.12) -- cycle ;
\draw  [fill={rgb, 255:red, 0; green, 0; blue, 0 }  ,fill opacity=1 ] (169.83,300.01) .. controls (170.38,299.92) and (170.89,300.29) .. (170.99,300.83) .. controls (171.08,301.38) and (170.71,301.89) .. (170.17,301.99) .. controls (169.62,302.08) and (169.11,301.71) .. (169.01,301.17) .. controls (168.92,300.62) and (169.29,300.11) .. (169.83,300.01) -- cycle ;
\draw  [fill={rgb, 255:red, 0; green, 0; blue, 0 }  ,fill opacity=1 ] (171.16,307.9) .. controls (171.71,307.81) and (172.22,308.18) .. (172.31,308.72) .. controls (172.41,309.27) and (172.04,309.78) .. (171.49,309.88) .. controls (170.95,309.97) and (170.43,309.6) .. (170.34,309.05) .. controls (170.25,308.51) and (170.62,307.99) .. (171.16,307.9) -- cycle ;
\draw  [fill={rgb, 255:red, 0; green, 0; blue, 0 }  ,fill opacity=1 ] (373.39,266.76) .. controls (373.9,266.53) and (374.49,266.76) .. (374.71,267.27) .. controls (374.93,267.78) and (374.7,268.37) .. (374.19,268.59) .. controls (373.69,268.81) and (373.1,268.58) .. (372.87,268.07) .. controls (372.65,267.57) and (372.88,266.98) .. (373.39,266.76) -- cycle ;
\draw  [fill={rgb, 255:red, 0; green, 0; blue, 0 }  ,fill opacity=1 ] (376.6,274.08) .. controls (377.1,273.86) and (377.69,274.09) .. (377.92,274.6) .. controls (378.14,275.1) and (377.91,275.69) .. (377.4,275.92) .. controls (376.9,276.14) and (376.31,275.91) .. (376.08,275.4) .. controls (375.86,274.9) and (376.09,274.31) .. (376.6,274.08) -- cycle ;
\draw  [fill={rgb, 255:red, 0; green, 0; blue, 0 }  ,fill opacity=1 ] (379.81,281.41) .. controls (380.31,281.19) and (380.9,281.42) .. (381.13,281.93) .. controls (381.35,282.43) and (381.12,283.02) .. (380.61,283.24) .. controls (380.1,283.47) and (379.51,283.24) .. (379.29,282.73) .. controls (379.07,282.22) and (379.3,281.63) .. (379.81,281.41) -- cycle ;

\draw (288,28.4) node [anchor=north west][inner sep=0.75pt]  [font=\footnotesize]  {$ \begin{array}{l}
|E|=n\\
|L|=0
\end{array}$};
\draw (429,138.4) node [anchor=north west][inner sep=0.75pt]  [font=\footnotesize]  {$ \begin{array}{l}
|E|=n\\
|L|=1
\end{array}$};
\draw (336,137.4) node [anchor=north west][inner sep=0.75pt]  [font=\footnotesize]  {$ \begin{array}{l}
|E|=n\\
|L|=1
\end{array}$};
\draw (122,140.4) node [anchor=north west][inner sep=0.75pt]  [font=\fontsize{0.68em}{0.81em}\selectfont]  {$ \begin{array}{l}
|E|=n-1\\
|L|=0
\end{array}$};
\draw (43,231.4) node [anchor=north west][inner sep=0.75pt]  [font=\fontsize{0.68em}{0.81em}\selectfont]  {$ \begin{array}{l}
|E|=n-2\\
|L|=0
\end{array}$};
\draw (130,230.4) node [anchor=north west][inner sep=0.75pt]  [font=\fontsize{0.68em}{0.81em}\selectfont]  {$ \begin{array}{l}
|E|=n-1\\
|L|=1
\end{array}$};
\draw (74,349.4) node [anchor=north west][inner sep=0.75pt]  [font=\fontsize{0.68em}{0.81em}\selectfont]  {$ \begin{array}{l}
|E|=0\\
|L|=|W_{1} \times W_{2} |
\end{array}$};
\draw (202,140.4) node [anchor=north west][inner sep=0.75pt]  [font=\fontsize{0.68em}{0.81em}\selectfont]  {$ \begin{array}{l}
|E|=n-1\\
|L|=0
\end{array}$};

\end{tikzpicture}
\end{center}

            \begin{itemize}
                \item[-] When the input models have different number of edges, the algorithm terminates without any recursion. The algorithm takes the space required in one instance of the function. The function defines two variables Found (Boolean), which take space other than the input. So, an instance of the function takes $\mathcal{O}(1)$ space.
                \item[-] When the two models have same number of edges, the recursive calls are made at lines 11, 19, 30 and 41. Suppose $n$ is the number of edges in the input models, then if the recursive call is made from line 11 or 19, the number of edges for those models is strictly less than $n$ (namely, $n-1$). The algorithm initially starts with $|L| = 0$. Whenever a recursive call is made from line number 30 or 41, $|L|$ strictly increases. Next it should be noted that the recursive call is not made from line 11 or 19 if the input model has no edges, and the recursive call is not made from line 30 or 41 if $|L| =  |W_1\times W_2|$. With these observations, we can bound the depth of recursion tree by $|R_1|\times |W_1\times W_2|$. This shows that the algorithm terminates.
               \end{itemize}
               With the above observations, we see that the depth of the recursion tree is bounded by $|R_1|\times |W_1\times W_2|$. Therefore, the space used by the algorithm is $s\times|R_1|\times |W_1\times W_2|$, where $s$ is the space used by one instance of the function \textsf{s-Bisimilar}. The function defines two variables Found (Boolean), which take space other than the input. So, once again, one instance of the function takes $\mathcal{O}(1)$ space.
           \end{proof}


\noindent We have just provided an algorithm for sabotage bisimulation. What about the algorithms for the other notions of bisimulation? We note here that the only differences in the definitions of the distinct notions of bisimulation are in conditions (4) and (5). We also observe that the function \textsf{s-Bisimilar} has 5 parts corresponding to the 5 conditions of bisimulation. For algorithm 1, the line numbers 8-15 and 16-23 correspond to the checks for the conditions 4 and 5, respectively. The corresponding algorithm for point sabotage bisimulation (d-bisimulation) is similar in nature. In particular, the loops in lines 8 and 10 should run for all $u_1 \in W_1$ and $u_2 \in W_2$, instead of the edges in the model. Further, in lines 11 and 19 where a recursive call should be made on the models 
$(\mathcal{M}^\prime_1 = (W_1\setminus\{u_1\},R_1\setminus\{(u_1,v_1), (y_1,u_1): v_1, y_1\in W_1\},V^\prime_1), u_1)$, where, for all propositional letter $p$, $V^\prime_1(p) = V_1(p)\setminus \{u_1\}$, and $(\mathcal{M}^\prime_2 = (W_2\setminus\{u_2\},R_2\setminus\{(u_2,v_2), (y_2,u_2): v_2, y_2\in W_2\},V^\prime_2), u_2)$, where, for all propositional letter $p$, $V^\prime_2(p) = V_2(p)\setminus \{u_2\}$, we need to have another check for $u_1\neq w_1$ and $u_2\neq w_2$ where input models are pointed at $w_1$ and $w_2$, respectively. This check has to be added before the recursion call. The detailed algorithm with a proof of its correctness is provided in Appendix B.

\medskip

\noindent For the generalized sabotage bisimulation algorithm, we need to add one extra condition before line 11 of algorithm 1 which checks whether $((W_1,R_1,V_1), u_1)$ is generalized sabotage bisimilar to $((W_2,R_2,V_2), u_2)$ and $((W_1,R_1,V_1), v_1)$ is  generalized sabotage bisimilar to $((W_2,R_2,V_2), v_2)$, where, $e_1=(u_1,v_1)$ and $e_2=(u_2,v_2)$. The same check should be added before line 19, corresponding to condition 5. For generalized point sabotage bisimulation, one more corresponding check is required, namely, $(\mathcal{M}^\prime_1, u_1)$ is generalized point 
sabotage bisimlar to $(\mathcal{M}^\prime_2, u_2)$. Once again, the detailed algorithms with a discussion on their correctness are provided in Appendix B. Finally, we note that the changes mentioned in all these algorithms only make more branching in the recursion tree but the depth is not affected. Since this depth affects the space taken, the space complexity of all these algorithms remain the same.

\section{Further remarks}

Till now, we have presented several existing logics of graph change concerning link deletion and point deletion games in graphs and studied the notion of model comparison or bisimulation for these logics from both logic as well as algorithmic points of view, and in process we have also shed some light into the complexity of these problems. We now provide some discussion on few emerging lines of continuation of this work. 

\subsection{ Connecting link deletion and point deletion}

Let us note that a link deletion involves the removal of an edge in a graph, whereas a point deletion involves the removal of a point together with the edges incident on it. Thus, in essence, point sabotage bisimulation algorithm emerges from the sabotage bisimulation algorithm where the recursive calls get modified to a certain extent. However, the model translations provided in \cite{lilei} give us natural ways to reduce one algorithm to the generalized version of the other. This would lead to the reductions in between the generalized versions as well.

\medskip

\noindent {\bf From sabotage bisimulation to generalized point saboage bisimulation}: Let $i$ be a new propositional letter. Given a relational model $\mathcal{M}_0 = (W_0,R_0,V_0)$, the model $\mathcal{F}(\mathcal{M}_0) = (W,R,V)$ is defined as follows: 

\begin{itemize}
\item[(a)] $W = W_0\cup W_i$ where $W_i = \{(w,v,i)\mid (w,v)\in R_0 \text{ and } w,v\in W_0\}$
\item[(b)] $R = \{(w,(w,v,i)), ((w,v,i),v) \mid (w,v)\in R_0\}$
\item[(c)] $V: \mathcal{P}\cup \{i\}\rightarrow W$ is a valuation function such that $V(p) = V_0(p)$ for $p\in \mathcal{P}$ and $V(i) = W_i$.
\end{itemize}

\noindent Given any two models $\mathcal{M}_1$ and $\mathcal{M}_2$, we can consider the models $\mathcal{F}(\mathcal{M}_1)$ and $\mathcal{F}(\mathcal{M}_2)$. Then, checking sabotage bisimulation between $\mathcal{M}_1$ and $\mathcal{M}_2$ would amount to checking generalized point sabotage bisimulation (with some minor modifications)  between $\mathcal{F}(\mathcal{M}_1)$ and $\mathcal{F}(\mathcal{M}_2)$, using the newly introduced propositional letter $i$ as the point deletion formula.

\medskip

\noindent {\bf From point sabotage bisimulation to generalized saboage bisimulation}: Let $j$ be a new propositional letter. Given a relational model  $\mathcal{M}_0 = (W_0,R_0,V_0)$, the model $\mathcal{G}(\mathcal{M}_0) = (W, R, V)$ is defined as follows: 

\begin{itemize}
\item[(a)] $W = W_0\cup \{w_j\}$
\item[(b)] $R = \{(u,v) \mid u R_j v \text{ and } v R_j w_j\}$, where $R_j = R_0\cup \{(w,w_j)\mid w\in W_0\}$ 
\item[(c)] $V: \mathcal{P}\cup \{j\}\rightarrow W$ is a valuation function such that $V(p) = V_0(p)$ for $p\in \mathcal{P}$ and $V(j) = \{w_j\}$.
\end{itemize}
\noindent Given any two models $\mathcal{M}_1$ and $\mathcal{M}_2$, we can consider the models $\mathcal{G}(\mathcal{M}_1)$ and $\mathcal{G}(\mathcal{M}_2)$. Then checking point sabotage bisimulation between $\mathcal{M}_1$ and $\mathcal{M}_2$ would amount to checking generalized sabotage bisimulation (with some minor modifications)  between $\mathcal{G}(\mathcal{M}_1)$ and $\mathcal{G}(\mathcal{M}_2)$. The main idea is that the removal of a point $w$ will amount to deleting the link between $w$ and $w_j$. 

\subsection{On complexity}

The complexity for checking whether given two pointed models are bisimilar, in basic modal logic, is known to be in polynomial time \cite{bisimulation}. What exactly makes the problem of s-bisimilarity more complex? The additional conditions (4) and (5) in the definition of s-bisimilarity, compared to that of basic modal logic bisimilarity, requires a function that assigns a sequence of non-repeating edges in one model to a sequence of non-repeating edges in other model. Formally, it requires a bijection $f:N(R_1)\to N(R_2)$ with $N(R)$ denoting the set of sequences of non-repeating edges from the edge relation $R$. The function $f$ should additionally satisfy the condition that any sequence of length $n$ is mapped to a sequences of length $n$, for every $n$. If $|R_1|=|R_2|=n$, then there are $2^{n^{2^n}}$ such functions. Given such a function, we need to check whether it satisfies the conditions for s-bisimilarity on top of the two models being basic modal bisimilar. This condition is what makes this problem of s-bisimiarity more complex. If we can show that every such function that satisfies the conditions for s-bisimilarity is generated by a function $g:R_1\to R_2$, then we believe that the complexity of s-bisimilarity drops to the class NP. To draw an analogy, deciding whether given two graphs are isomorphic is in NP, but finding the isomorphism mapping may be more complex. This is equivalent to say that given a small (with number of elements bounded by a polynomial in the size of the input models) candidate generator of the relation s-bisimilar, it may be efficient to check whether such a candidate can be extended to a full s-bisimilar relation.

\medskip

\noindent {\bf Connection with local sabotage bisimulation}:
It has already been shown that if the two given pointed models (each with finite branching) are s-bisimilar, then it implies that they are local s-bisimilar \cite{sab-logic}. This gives us a straight forward reduction from local s-bisimilarity to s-bisimilarity. Therefore, the complexity of deciding whether given two pointed models are local s-bisimilar cannot be more than the complexity of s-bisimilarity. An interesting question in this context is whether the complexity of local s-bisimilarity is strictly less than the complexity of s-bisimilarity. A known result in the same context is that the complexity of model checking for local s-bisimilarity is stricltly less than that of sabotage modal logic (P-time vs PSPACE-complete). If it indeed turns out that complexity of s-bisimilarity is in NP, then the structure of the candidate generator, which is equivalent to giving bijection between edges of both models, would also work for local s-bisimilarity and we suspect that in this case, both the problems will have same complexity, namely, NP. On the other hand, if it turns out that no such small certificate (candidate generator) for s-bisimilarity exists, then we suspect that deciding whether given two pointed models are local s-bisimilar maybe strictly in lower complexity class, compared to s-bisimilarity.

\subsection{Revisiting games on graphs}

We started off our discussion with games played on graphs and we would like to end on the same note. What does it mean to have a bisimulation between two game graphs with respect to two points on those two graphs? Evidently, whatever moves a player can make in one game, the same kind of moves can be made in the other game as well. Moreover, an alternation of the basic modality with the sabotage or removal modality would describe a play in the game graphs with link deletion or point deletion, respectively. From the strategic viewpoint, the age-old copy strategy might be a relevant strategy to play on bisimilar game graphs. In fact, checking bisimilarity between different game graphs can be considered as a first step towards considering game-strategy equivalences between these games constituting structural changes in the underlying graphs. 

\newpage

\nocite{*}
\bibliographystyle{unsrt}  
\bibliography{template}


    

\newpage 

\section*{Appendix A: Characterizing formulas for bisimulations}

\subsection*{Point sabotage bisimulation (d-bisimulation)}

Suppose there are $n$ worlds in a model $M=(W,R,V)$, let $M_{[e_1,\ ...\ , e_k]}$ denote the new model after we delete the worlds $e_1,\ ...\ , e_k$. Let $Q_{(e_1,...,e_k)}$ denote $(e_1,\ ...\ ,e_k)\in W^{k}\And e_i\not=e_j,\ \text{if}\ i\not=j$. When $1\leq k <n$, let  $H_k(M)$ denote 

$$ \bigwedge_{Q_{(e_1,...,e_k)}} {\left\langle -\right\rangle }^{k}E(M_{[e_1,\ ...\ , e_k]})\ \bigwedge {\left[ -\right] }^{k}\bigvee_{Q_{(e_1,...,e_k)}} E(M_{[e_1,\ ...\ , e_k]})$$
Let $H_{n}(M)$ be $\neg{\left\langle -\right\rangle}^{n}\top$, $H(M)$ be $E(M)\wedge\bigwedge_{1\leq i\leq n}H_i(M)$. Intuitively, $Q_{(e_1,...,e_k)}$ means the class of point sequences which consist of $k$ points different from each other. If $H(M)$ is valid on some model, which means the new  model after some point-deleting sequence keeps similar to the corresponding model  after some point-deleting sequence. Especially, the case that the point-deleting sequence is empty, is the situation on basic modal logic. The following theorem provides a connection between d-bisimulation and the formula described above. Let the symbol $\underline{\leftrightarrow}_d$ indicate the existence of a d-bisimulation between two pointed models.
\begin{theorem}
	For any two pointed models $(M,s)=(W_1,R_1,V_1,s)$ and $(N,t)= (W_2,R_2,V_2,t)$, the following are equivalent:
	\begin{itemize}
	\item[$(a)$] $(M,s)\underline{\leftrightarrow}_d (N,t)$
	\item[$(b)$] $(N,t)$ can be expanded to a model $(N^\prime,t)$ for $H(M)$ such that $p_s$ holds at world $t$.
	\end{itemize} 
\end{theorem}

\begin{proof}
	$[(a)\Longrightarrow(b)]$ Define $N^\prime=(W_2,R_2,V^\prime_2)$, where $V_{2}^{\prime}(p)= V_2 (p)$ for any proposition letter in the initial language,   $V_{2}^{\prime}(p_x)=\{u|(M,x)\underline{\leftrightarrow}_d (N,u) \}$ for $x\in W_1$. We have to prove  
	$N^\prime\vDash H(M)\And N^\prime, t\vDash p_{s}$. Since  $(M,s)\underline{\leftrightarrow}_d (N,t)$, then $t\in V_{2}^{\prime}(p_s)$, which means $p_s$ holds at world $t$. We have to prove $N^\prime\vDash H(M)$. Since $H(M)$ is $E(M)\wedge\bigwedge_{1\leq i\leq n}H_i(M)$. Firstly, we prove that  $N^\prime\nvDash E(M)$ by contradiction.
	
	Since $E(M)$ is the conjuction of all statements $p_x\rightarrow AT_x\wedge env(M,x)$, where $env(M,x)$ is the formula $\bigwedge_{y\in \{y|R_1 xy\}}\Diamond p_{y} \wedge \Box\bigvee_{y\in \{y|R_1 xy\}} p_{y}$.
	Suppose $N^\prime\nvDash E(M)$, then there exists $e$ in $N^\prime$ such that $N^\prime, e \nvDash p_y\rightarrow AT_y \wedge env(M,y)$ for some $y\in W_1$, then $N^\prime, e \vDash p_y$ and $N^\prime, e \nvDash AT_y \wedge env(M,y)$. Since $N^\prime, e \vDash p_y$, then $(M,y)\underline{\leftrightarrow}_d (N,e)$, then $(M,y)\vDash p$ iff $N,e\nvDash p$ for any proposition letter $p$ in the initial language.
	\begin{itemize}
		\item [$\bullet$] If $N^\prime, e \nvDash AT_y$, then there exists $p$ in the initial language such that $(M,y)\vDash p$ iff $N^\prime,e\nvDash p$ iff $N,e\nvDash p$, contradiction.
		\item [$\bullet$] If $N^\prime, e \nvDash  env(M,y)$, then $N^\prime, e \nvDash \bigwedge_{z\in \{y|R_{1} yz\}}\Diamond p_{z} \wedge \Box\bigvee_{z\in \{y|R_{1} yz\}} p_{z}$.
		\begin{itemize}
			\item[$-$] If $N^\prime, e \nvDash \bigwedge_{z\in \{y|R_{1} yz\}}\Diamond p_{z} $, then  $N^\prime, e \nvDash \Diamond p_z$ for some $z$ with $R_{1}yz$, since $(M,y)\underline{\leftrightarrow}_d (N,e)$, then there exists $z^\prime$ with $R_2e{z}^\prime$ and $(M,z)\underline{\leftrightarrow}_d (N,z^\prime)$, then $z^\prime\in V_{2^{\prime}}(p_z)$, thus $N,z^\prime\vDash p_z$, then we have $N,e\vDash\Diamond p_z$, contradiction.
			\item[$-$] If $N^\prime, e \nvDash \Box\bigvee_{z\in \{z|R_{1} yz\}} p_{z}$, then there exists world $e^\prime$ with $R_{2}ee^\prime$, such that $N^\prime, e^\prime \vDash \bigwedge_{z\in \{y|R_{1} yz\}}\neg p_z$. since $(M,y)\underline{\leftrightarrow}_d (N,e)$, then there exists $y^\prime$ with $R_{1}yy^\prime$ such that $(M,y^\prime)\underline{\leftrightarrow}_d (N,e^\prime)$, then $p_{y^\prime}$ holds at world $e^\prime$, contradiction.
		\end{itemize}
	\end{itemize}
Then we prove $N^\prime\vDash\bigwedge_{1\leq i\leq n}H_i(M)$ directly. For any  point-deleting sequence $(d_1,...,d_l)$ satisfying  $Q_{(d_1,...,d_l)}$ from  $M$, because of the $Zig_{\left\langle -\right\rangle }$ item of d-bisimulation, there exists a point-deleting sequence $(j_1,...,j_l)$ satisfying  $Q_{(j_1,...,j_l)}$ from  $N$, such that  $(M_{[d_1,...,d_l]},s)\underline{\leftrightarrow}_d (N_{[j_1,...,j_l]},t)$. Similar to the first part, we have that $(N_{[j_1,...,j_l]},t)$ can be expanded to a model $({N_{[j_1,...,j_l]}}^\prime,t)$ for $E(M_{[d_1,...,d_l]})$, since ${N_{[j_1,...,j_l]}}^\prime= N_{[j_1,...,j_l]}^\prime$, then $ N^\prime\vDash {\left\langle  - \right\rangle }^{l}E(M_{[(d_1,...,d_l)]})$, then $ N^\prime\vDash \bigwedge_{P_{(i_1,...,i_l)}} {\left\langle  - \right\rangle }^{l}E(M_{[i_1,\ ...\ , i_l]})$. Since for any point-deleting sequence $(j_1,...,j_l)$ satisfying  $Q_{(j_1,...,j_l)}$ from  $N$, we have $ N_{[j_1,...,j_l]}^\prime\vDash \bigvee_{P_{(i_1,...,i_l)}} E(M_{[i_1,\ ...\ , i_l]})$, then $ N^\prime\vDash{ [-]}^{l}\bigvee_{P_{(i_1,...,i_l)}} E(M_{[i_1,\ ...\ , i_l]})$. And since model $N$ has the same number of edges with model $M$, Thus $N^\prime\vDash \neg {\left\langle -  \right\rangle }^{n}\top$, then $N^\prime\vDash H(M)$.

\smallskip

	$[(a)\Longleftarrow(b)]$ We prove this direction by contradiction. If $(M,s)\cancel{$\underline\leftrightarrow$}_d (N,t)$, by the definition of d-bisimulation, there are the following cases:
	\begin{itemize}
		\item[$(1)$]  $ s\in V_{1}(p)$ iff $t\notin V_{2}(p)$ for some proposition letter $p$.
		\item[$(2)$] there exists some world $s_1$ with $R_1 ss_1$, there is no $t_1$ with $R_2 tt_1$.
		\item[$(3)$] there exists some world $t_1$ with $R_2 tt_1$, there is no $s_1$ with $R_1 ss_1$.
		\item[$(4)$] there is a world $s_1$ with $R_1 ss_1$, for any world $t_1$ with $R_2 tt_1$, $(M,s_1)\cancel{$\underline\leftrightarrow$}_d (N,t_1)$.
		\item[$(5)$] there is a world $t_1$ with $R_2 tt_1$, for any world $s_1$ with $R_1 ss_1$, $(M,s_1)\cancel{$\underline\leftrightarrow$}_d (N,t_1)$.
		\item[$(6)$] if we delete a world in the model $M$, there is no other world different from $t$ in the model $N$.
		\item[$(7)$] if we delete a world in the model $N$, there is no other world different from $s$ in the model $M$.
		\item[$(8)$] if we delete a world $d_1$ in the model $M$, for any world $j_1$ in the model $N$, $(M_{[d_1]},s)\cancel{$\underline\leftrightarrow$}_d (N_{[j_1]},t)$.
		\item[$(9)$] if we  delete a world $j_1$ in the model $N$, for any world $d_1$ in the model $M$, $(M_{[d_1]},s)\cancel{$\underline\leftrightarrow$}_d (N_{[j_1]},t)$.
	\end{itemize}
	For the cases $(4),(5),(7),(8)$, we can follow the definition of d-bisimulation, and divide them into more cases. Along this way, we can find that all the cases are excluded by the following lemmas. So it is sufficient to prove Lemma $12$ and $13$ given below.
\end{proof}	
	
	\begin{lemma}
		For any point-deleting sequence $(d_1,...,d_l)$ satisfying $Q_{(d_1,...,d_l)}$ from  $M$, where $0\leq l< n$, if there is a path $s_0\rightarrow s_1\rightarrow\ ...\ \rightarrow s_k$ in $M_{[d_1,...,d_l]}$, then there exists a point-deleting sequence $(j_1,...,j_l)$ from  $N$ and a path  $t_0\rightarrow t_1\rightarrow\ ...\ \rightarrow t_k$ in $N_{[j_1,...,j_l]}$ such that $s_i\in V_{1}(p)$ iff $t_i\in V_{2}(p)$ for each proposition letter $p$ in the initial language, where $s_0$ is reachable from $s$ in $M$ , $t_0$ is reachable from $t$ in $N$, $i\in [0,k]$. 		
	\end{lemma}
	
	\begin{proof}
		For any path $s_0\rightarrow s_1\rightarrow\ ...\ \rightarrow s_k$ in $M_{[d_1,...,d_l]}$, since $ N^\prime\vDash \bigwedge_{Q_{(i_1,...,i_l)}} {\left\langle - \right\rangle }^{l}E(M_{[i_1,\ ...\ , i_l]})$, then $ N^\prime\vDash {\left\langle - \right\rangle }^{l}E(M_{[d_1,...,d_l]})$, thus  $ N^\prime_{[j_1,...,j_l]}\vDash E(M_{[d_1,...,d_l]})$ for some point-deleting sequence $(j_1,...,j_l)$ from  $N^\prime$.	We take $s_0$ satisfying $N^\prime,t_0\vDash p_{s_0}$, since $N^\prime_{[j_1,...,j_l]}, t_0\vDash p_{s_0}\rightarrow AT_{s_0} \wedge env(M_{[d_1,...,d_l]},s)$, then $N^\prime,t_0\vDash AT_{s_0}$. Suppose that we have select $t_i(i<k)$ satisfying $N_{[j_1,...,j_l]}^\prime,t_i\vDash  p_{s_i}$,  we prove that there exists $t_{i+1}$ such that $N^\prime_{[j_1,...,j_l]}, t_{i+1}\vDash p_{s_{i+1}}$ and $R_{2}t_{i}t_{i+1}$. Since $N^\prime_{[j_1,...,j_l]}, t_i\vDash p_{s_i}\rightarrow AT_{s_i} \wedge env(M_{[d_1,...,d_l]},s_i)$, then $N^\prime_{[j_1,...,j_l]},t_i\vDash AT_{s_i}$, and $N^\prime_{[j_1,...,j_l]},t_i\vDash env(M_{[d_1,...,d_l]},s_i)$. Thus we have $N^\prime_{[j_1,...,j_l]},t_i\vDash \Diamond p_{s_{i+1}}$, then  there exists $e$ with $R_{2} t_{i}e$ such that $N^\prime_{[j_1,...,j_l]},e\vDash p_{s_{i+1}}$, since 	$N^\prime_{[j_1,...,j_l]}, e\vDash p_{s_{i+1}}\rightarrow AT_{s_{i+1}} \wedge env(M_{[d_1,...,d_l]},s_{i+1})$, then 	$N^\prime_{[j_1,...,j_l]}, e\vDash AT_{s_{i+1}}$. Let $t_{i+1}$ be $e$, then we finish our lemma.
	\end{proof}
	
	\begin{lemma}
		For any point-deleting sequence $(j_1,...,j_l)$ satisfying $Q_{(j_1,...,j_l)}$ from  $N$, where $0\leq l< n$, if there is a path $t_0\rightarrow t_1\rightarrow\ ...\ \rightarrow t_k$ in $N_{[j_1,...,j_l]}$ , then there exists a point-deleting sequence $(d_1,...,d_l)$ from  $M$ and a path $s_0\rightarrow s_1\rightarrow\ ...\ \rightarrow s_k$ in $M_{[d_1,...,d_l]}$  such that $s_i\in V_{1}(p)$ iff $t_i\in V_{2}(p)$ for each proposition letter $p$ in the initial language, where $t_0$ is reachable from $t$ in $N$, $s_0$ is reachable from $s$ in $M$, $i\in [0,k]$.
	\end{lemma}
	
	\begin{proof}
		Since $ N^\prime\vDash {[-]}^{l}\bigvee_{P_{(i_1,...,i_l)}} E(M_{[i_1,\ ...\ , i_l]})$, then for any point-deleting sequence $(j_1,...,j_l)$ from  $N^\prime$,  $N_{[j_1,...,j_l]}^\prime \vDash E(M_{[d_1,...,d_l]})$ for some point-deleting sequence $(d_1,...,d_l)$ from  $M$. For any path $t_0\rightarrow t_1\rightarrow\ ...\ \rightarrow t_k$ in $N_{[j_1,...,j_l]}$, we select $s_0$ satisfing $N_{[j_1,...,j_l]}^\prime,t_0\vDash p_{s_{0}}$, since 
		$N^\prime_{[j_1,...,j_l]}, t_0\vDash p_{s_0}\rightarrow AT_{s_0} \wedge env(M_{[d_1,...,d_l]},s_0)$, then $N^\prime_{[j_1,...,j_l]},t_0\vDash AT_{s_0}$.  Suppose that we have select $s_i(i<k)$ satisfying $N_{[j_1,...,j_l]}^\prime,t_i\vDash  p_{s_i}$, we prove that there exists $s_{i+1}$ such that $N^\prime, t_{i+1}\vDash p_{s_{i+1}}$ and $R_{1}s_{i}s_{i+1}$. Since $N^\prime_{[j_1,...,j_l]}, t_i\vDash p_{s_i}\rightarrow AT_{s_i} \wedge env(M_{[d_1,...,d_l]},s_i)$, then $N^\prime_{[j_1,...,j_l]},t_i\vDash AT_{s_i}$, and $N^\prime_{[j_1,...,j_l]},t_i\vDash env(M_{[d_1,...,d_l]},s_i)$. If $env(M_{[d_1,...,d_l]},s_i)$ is $\top\wedge\Box\bot$(when $s_i$ has no successor in $M_{[d_1,...,d_l]}$),  then $N^\prime_{[j_1,...,j_l]},t_i\vDash\Box\bot$, it's impossible since $t_{i+1}$ is a successor of world $t_i$ in $N^\prime_{[j_1,...,j_l]}$, then there exists $z$ with $R_{1}s_{i}z$ such that $N^\prime_{[j_1,...,j_l]}, t_{i+1}\vDash p_{z}$, since 
		$N^\prime_{[j_1,...,j_l]}, t_{i+1}\vDash p_{z}\rightarrow AT_{z} \wedge env(M_{[d_1,...,d_l]},z)$, then $N^\prime_{[j_1,...,j_l]}, t_{i+1}\vDash  AT_{z}$. Let $s_{i+1}$ be  $z$, then we finish our proof.
	\end{proof}

\subsection*{General sabotage bisimulation (g-bisimulation)}

We move on to general sabotage modal logic, to characterize g-bisimulation.
Let the edge $i=(i(1),i(2))$, where $i(1),i(2)\in W$. For $1\leq k\leq  n$, let  $K_k(M)$ denote: 
$$ \bigwedge_{P_{(i_1,...,i_k)}} \blacklozenge^{p_{i_{1}(1)}}_{p_{i_{1}(2)}}...\ \blacklozenge^{p_{i_{k}(1)}}_{p_{i_{k}(2)}}E(M_{(i_1,\ ...\ , i_k)})\ \bigwedge \blacksquare^{p_{i_{1}(1)}}_{p_{i_{1}(2)}}...\ \blacksquare^{p_{i_{k}(1)}}_{p_{i_{k}(2)}}\bigvee_{P_{(i_1,...,i_k)}} E(M_{(i_1,\ ...\ , i_k)})$$
Let $K_{n+1}(M)$ be $\neg ({\blacklozenge^{\top}_{\top}})^{n+1}\top$, $K(M)$ be $E(M)\wedge\bigwedge_{1\leq i\leq n+1}K_i(M)$. Intuitively,
since we add new proposition letter $p_w$ if $w$ is a world in $M$, an edge $i$ can be marked by two new proposition letters $p_{i(1)},p_{i(2)}$. We modify the formula $G(M)$ as $K(M)$ to fit g-bisimulation. Let the symbol $\underline{\leftrightarrow}_g$ indicate the existence of g-bisimulation between two pointed models.
\begin{theorem}
	For any two pointed models $(M,s)=(W_1,R_1,V_1,s)$ and $(N,t)= (W_2,R_2,V_2,t)$, the following are equivalent:
	\begin{itemize}
		\item[$(a)$] $(M,s)\underline{\leftrightarrow}_g (N,t)$
		\item[$(b)$] $(N,t)$ can be expanded to a model $(N^\prime,t)$ for $K(M)$ such that $p_s$ holds at world $t$.
	\end{itemize} 
\end{theorem}
\begin{proof}
	$[(a)\Longrightarrow(b)]$  Define $N^\prime=(W_2,R_2,V^\prime_2)$, where $V_{2}^{\prime}(p)= V_2 (p)$ for any proposition letter in the initial language,   $V_{2}^{\prime}(p_x)=\{u|(M,x)\underline{\leftrightarrow}_g (N,u) \}$ for $x\in W_1$. We have to prove 
	$N^\prime\vDash K(M)\And N^\prime, t\vDash p_{s}$. Since  $(M,s)\underline{\leftrightarrow}_g (N,t)$, then $t\in V_{2}^{\prime}(p_s)$, which means $p_s$ holds at world $t$. We have to prove $N^\prime\vDash K(M)$. Since $K(M)$ is $E(M)\wedge\bigwedge_{1\leq i\leq n+1}K_i(M)$. Firstly, we prove that  $N^\prime\nvDash E(M)$ by contradiction.
	
	Since $E(M)$ is the conjuction of all statements $p_x\rightarrow AT_x\wedge env(M,x)$, where $env(M,x)$ is the formula $\bigwedge_{y\in \{y|R_1 xy\}}\Diamond p_{y} \wedge \Box\bigvee_{y\in \{y|R_1 xy\}} p_{y}$.
	Suppose $N^\prime\nvDash E(M)$, then there exists $e$ in $N^\prime$ such that $N^\prime, e \nvDash p_y\rightarrow AT_y \wedge env(M,y)$ for some $y\in W_1$, then $N^\prime, e \vDash p_y$ and $N^\prime, e \nvDash AT_y \wedge env(M,y)$. Since $N^\prime, e \vDash p_y$, then $(M,y)\underline{\leftrightarrow}_g (N,e)$, then $(M,y)\vDash p$ iff $N,e\nvDash p$ for any proposition letter $p$ in the initial language.
	\begin{itemize}
		\item [$\bullet$] If $N^\prime, e \nvDash AT_y$, then there exists $p$ in the initial language such that $(M,y)\vDash p$ iff $N^\prime,e\nvDash p$ iff $N,e\nvDash p$, contradiction.
		\item [$\bullet$] If $N^\prime, e \nvDash  env(M,y)$, then $N^\prime, e \nvDash \bigwedge_{z\in \{y|R_{1} yz\}}\Diamond p_{z} \wedge \Box\bigvee_{z\in \{y|R_{1} yz\}} p_{z}$.
		\begin{itemize}
			\item[$-$] If $N^\prime, e \nvDash \bigwedge_{z\in \{y|R_{1} yz\}}\Diamond p_{z} $, then  $N^\prime, e \nvDash \Diamond p_z$ for some $z$ with $R_{1}yz$, since $(M,y)\underline{\leftrightarrow}_g (N,e)$, then there exists $z^\prime$ with $R_2e{z}^\prime$ and $(M,z)\underline{\leftrightarrow}_g (N,z^\prime)$, then $z^\prime\in V_{2^{\prime}}(p_z)$, thus $N,z^\prime\vDash p_z$, then we have $N,e\vDash\Diamond p_z$, contradiction.
			\item[$-$] If $N^\prime, e \nvDash \Box\bigvee_{z\in \{z|R_{1} yz\}} p_{z}$, then there exists world $e^\prime$ with $R_{2}ee^\prime$, such that $N^\prime, e^\prime \vDash \bigwedge_{z\in \{y|R_{1} yz\}}\neg p_z$. since $(M,y)\underline{\leftrightarrow}_g (N,e)$, then there exists $y^\prime$ with $R_{1}yy^\prime$ such that $(M,y^\prime)\underline{\leftrightarrow}_g (N,e^\prime)$, then $p_{y^\prime}$ holds at world $e^\prime$, contradiction.
		\end{itemize}
	\end{itemize}
	Then we prove $N^\prime\vDash\bigwedge_{1\leq i\leq n+1}K_i(M)$ directly. For any  link-cutting sequence $(d_1,...,d_l)$ satisfying $P_{(d_1,...,d_l)}$ from  $M$, because of the $Zig_{\blacklozenge^{\psi}_{\psi}}$ item of g-bisimulation, there exists a link-cutting sequence $(j_1,...,j_l)$ satisfing $P_{(j_1,...,j_l)}$ from  $N$, such that for $0\leq m<l$, $(M_{(d_1,...,d_{m})},d_{m+1}(1))\underline{\leftrightarrow}_g (N_{(j_1,...,j_{m})},j_{m+1}(1))$, $(M_{(d_1,...,d_{m})},d_{m+1}(2))\underline{\leftrightarrow}$ 
	
	\noindent $(N_{(j_1,...,j_{m})},j_{m+1}(2))$ and  $(M_{(d_1,...,d_{m+1})},s)\underline{\leftrightarrow}_g (N_{(j_1,...,j_l{m+1})},t)$. Similarly with the first part, we have that $(N_{(j_1,...,j_l)},$
	
	\noindent $t)$ can be expanded to a model $({N_{(j_1,...,j_l)}}^\prime,t)$ for $E(M_{(d_1,...,d_l)})$, since ${N_{(j_1,...,j_l)}}^\prime= N_{(j_1,...,j_l)}^\prime$, then $ N^\prime\vDash \blacklozenge^{p_{d_{1}(1)}}_{p_{d_{1}(2)}}\ ...\ \blacklozenge^{p_{d_{l}(1)}}_{p_{d_{l}(2)}}$
	
	\noindent $E(M_{(d_1,...,d_l)})$, thus $ N^\prime\vDash \bigwedge_{P_{(i_1,...,i_l)}} \blacklozenge^{p_{i_{1}(1)}}_{p_{i_{1}(2)}}\ ...\ \blacklozenge^{p_{i_{l}(1)}}_{p_{i_{l}(2)}}E(M_{(i_1,\ ...\ , i_l)})$. Since for any link-cutting sequence $(j_1,...,j_l)$ satisfing $P_{(j_1,...,j_l)}$, we have $ N_{(j_1,...,j_l)}^\prime\vDash \bigvee_{P_{(i_1,...,i_l)}} E(M_{(i_1,\ ...\ , i_l)})$, then $ N^\prime\vDash \blacksquare^{p_{i_{1}(1)}}_{p_{i_{1}(2)}}\ ...\ \blacksquare^{p_{i_{l}(1)}}_{p_{i_{l}(2)}}\bigvee_{P_{(i_1,...,i_l)}} E(M_{(i_1,\ ...\ , i_l)})$. And since model $N$ has the same number of edges with model $M$, Thus $N^\prime\vDash \neg ({\blacklozenge^{\top}_{\top}})^{n+1}\top$, then $N^\prime\vDash K(M)$.
	
	\smallskip
	
	$[(a)\Longleftarrow(b)]$ To prove this direction it is sufficient to prove Lemmas $15$ and $16$ given below. Lemma $15$ is for the $Zig$ item for  $\Diamond\And\blacklozenge$, while lemma $16$ is for the $Zag$ item for  $\Diamond\And\blacklozenge$. 
\end{proof}

\begin{lemma}
	For any link-cutting sequence $(d_1,...,d_l)$ satisfying $P_{(d_1,...,d_l)}$ from  $M$, where $0\leq l\leq n$, if there is a  path $s_0\rightarrow s_1\rightarrow\ ...\ \rightarrow s_k$ in $M_{(d_1,...,d_l)}$, then there exists a link-cutting sequence $(j_1,...,j_l)$ satisfying $P_{(j_1,...,j_l)}$ from  $N$, $p_{d_m(1)}$ holds at $j_m(1)$, $p_{d_m(2)}$ holds at $j_m(2)$ for $1\leq m\leq l$, and a path  $t_0\rightarrow t_1\rightarrow\ ...\ \rightarrow t_k$ in $N_{(j_1,...,j_l)}$ such that $s_i\in V_{1}(p)$ iff $t_i\in V_{2}(p)$ for each proposition letter $p$ in the initial language, where $s_0$ is reachable from $s$ in $M$ , $t_0$ is reachable from $t$ in $N$, $i\in [0,k]$.
\end{lemma}
\begin{proof}
	For any path $s_0\rightarrow s_1\rightarrow\ ...\ \rightarrow s_k$ in $M_{(d_1,...,d_l)}$, since $ N^\prime\vDash \bigwedge_{P_{(i_1,...,i_l)}} \blacklozenge^{p_{i_{1}(1)}}_{p_{i_{1}(2)}}\ ...\ \blacklozenge^{p_{i_{l}(1)}}_{p_{i_{l}(2)}}E(M_{(i_1,\ ...\ , i_l)})$, then $ N^\prime\vDash \blacklozenge^{p_{d_{1}(1)}}_{p_{d_{1}(2)}}\ ...\ \blacklozenge^{p_{d_{l}(1)}}_{p_{d_{l}(2)}}E(M_{(d_1,\ ...\ , d_l)})$, thus  $ N^\prime_{(j_1,...,j_l)}\vDash E(M_{(d_1,\ ...\ , d_l)})$ for some link-cutting sequence $(j_1,...,j_l)$ from  $N^\prime$, and $p_{d_m(1)}$ holds at $j_m(1)$, $p_{d_m(2)}$ holds at $j_m(2)$ for $1\leq m\leq l$. We take $s_0$ satisfying $N^\prime,t_0\vDash p_{s_0}$, since $N^\prime_{(j_1,...,j_l)}, t_0\vDash p_{s_0}\rightarrow AT_{s_0} \wedge env(M_{(d_1,...,d_l)},s_0)$, then $N^\prime,t_0\vDash AT_{s_0}$. Suppose that we have select $t_i(i<k)$ satisfying $N_{(j_1,...,j_l)}^\prime,t_i\vDash  p_{s_i}$, we prove that there exists $t_{i+1}$ such that $N^\prime_{(j_1,...,j_l)}, t_{i+1}\vDash p_{s_{i+1}}$ and $R_{2}t_{i}t_{i+1}$. Since $N^\prime_{(j_1,...,j_l)}, t_i\vDash p_{s_i}\rightarrow AT_{s_i} \wedge env(M_{(d_1,...,d_l)},s_i)$, then $N^\prime_{(j_1,...,j_l)},t_i\vDash AT_{s_i}$, and $N^\prime_{(j_1,...,j_l)},t_i\vDash env(M_{(d_1,...,d_l)},s_i)$. Thus we have $N^\prime_{(j_1,...,j_l)},t_i\vDash \Diamond p_{s_{i+1}}$, then  there exists $e$ with $R_{2} t_{i}e$ such that $N^\prime_{(j_1,...,j_l)},e\vDash p_{s_{i+1}}$, since 	$N^\prime_{(j_1,...,j_l)}, e\vDash p_{s_{i+1}}\rightarrow AT_{s_{i+1}} \wedge env(M_{(d_1,...,d_l)},s_{i+1})$, then 	$N^\prime_{(j_1,...,j_l)}, e\vDash AT_{s_{i+1}}$. Let $t_{i+1}$ be $e$, then we finish our lemma.	
\end{proof}

\begin{lemma}
	For any link-cutting sequence $(j_1,...,j_l)$ satisfying $P_{(j_1,...,j_l)}$ from  $N$, where $0\leq l\leq n$, if there is a path $t_0\rightarrow t_1\rightarrow\ ...\ \rightarrow t_k$ in $N_{(j_1,...,j_l)}$ , then there exists a link-cutting sequence $(d_1,...,d_l)$  satisfying $P_{(d_1,...,d_l)}$ from $M$,$p_{d_m(1)}$ holds at $j_m(1)$, $p_{d_m(2)}$ holds at $j_m(2)$ for $1\leq m\leq l$, and a path $s_0\rightarrow s_1\rightarrow\ ...\ \rightarrow s_k$ in $M_{(d_1,...,d_l)}$  such that $s_i\in V_{1}(p)$ iff $t_i\in V_{2}(p)$ for each proposition letter $p$ in the initial language, where $t_0$ is reachable from $t$ in $N$, $s_0$ is reachable from $s$ in $M$, $i\in [0,k]$.
\end{lemma}

\begin{proof}	
	Since $ N^\prime\vDash\bigwedge_{P_{(i_1,...,i_n)}} \blacklozenge^{p_{i_{1}(1)}}_{p_{i_{1}(2)}}...\ \blacklozenge^{p_{i_{n}(1)}}_{p_{i_{n}(2)}}E(M_{(i_1,\ ...\ , i_n)})$, then for any world $v$ in $N^\prime$, it must be statisfied by $p_w$ for some world $v$ in $M$. Since $N^\prime\vDash\blacksquare^{p_{i_{1}(1)}}_{p_{i_{1}(2)}}\ ...\ \blacksquare^{p_{i_{l}(1)}}_{p_{i_{l}(2)}}\bigvee_{P_{(i_1,...,i_l)}} E(M_{(i_1,\ ...\ , i_l)})$, then for any link-cutting sequence $(j_1,...,j_l)$ from  $N$,  $N_{(j_1,...,j_l)}^\prime \vDash E(M_{(d_1,...,d_l)})$ for some link-cutting sequence $(d_1,...,d_l)$ from  $M$ satisfying $p_{d_m(1)}$ holds at $j_m(1)$, $p_{d_m(2)}$ holds at $j_m(2)$ for $1\leq m\leq l$. For any path $t_0\rightarrow t_1\rightarrow\ ...\ \rightarrow t_k$ in $N_{(j_1,...,j_l)}$, we select $s_0$ satisfing $N_{(j_1,...,j_l)}^\prime,t_0\vDash p_{s_{0}}$, since 
	$N^\prime_{(j_1,...,j_l)}, t_0\vDash p_{s_0}\rightarrow AT_{s_0} \wedge env(M_{(d_1,...,d_l)},s_0)$, then $N^\prime_{(j_1,...,j_l)},t_0\vDash AT_{s_0}$.  Suppose that we have select $s_i(i<k)$ satisfying $N_{(j_1,...,j_l)}^\prime,t_i\vDash  p_{s_i}$, we prove that there exists $s_{i+1}$ such that $N^\prime, t_{i+1}\vDash p_{s_{i+1}}$ and $R_{1}s_{i}s_{i+1}$. Since $N^\prime_{(j_1,...,j_l)}, t_i\vDash p_{s_i}\rightarrow AT_{s_i} \wedge env(M_{(d_1,...,d_l)},s_i)$, then $N^\prime_{(j_1,...,j_l)},t_i\vDash AT_{s_i}$, and $N^\prime_{(j_1,...,j_l)},t_i\vDash env(M_{(d_1,...,d_l)},s_i)$. If $env(M_{(d_1,...,d_l)},s_i)$ is $\top\wedge\Box\bot$(when $s_i$ has no successor in $M_{(d_1,...,d_l)}$),  then $N^\prime_{(j_1,...,j_l)},t_i\vDash\Box\bot$, it's impossible since $t_{i+1}$ is a successor of world $t_i$ in $N^\prime_{(j_1,...,j_l)}$, then there exists $z$ with $R_{1}s_{i}z$ such that $N^\prime_{(j_1,...,j_l)}, t_{i+1}\vDash p_{z}$, since 
	$N^\prime_{(j_1,...,j_l)}, t_{i+1}\vDash p_{z}\rightarrow AT_{z} \wedge env(M_{(d_1,...,d_l)},z)$, then $N^\prime_{(j_1,...,j_l)}, t_{i+1}\vDash  AT_{z}$. Let $s_{i+1}$ be  $z$, then we finish our proof.
\end{proof}

\subsection*{General point sabotage bisimulation (r-bisimulation)}

Now we prove the similar result for r-bisimulation. When $1\leq k <n$, let  $L_k(M)$ denote 

$$ \bigwedge_{Q_{(e_1,...,e_k)}} {\left\langle -p_{e_1}\right\rangle }...\ {\left\langle -p_{e_k}\right\rangle }E(M_{[e_1,\ ...\ , e_k]})\ \bigwedge {\left[ -p_{e_1}\right] }...\ {\left[ -p_{e_k}\right] }\bigvee_{Q_{(e_1,...,e_k)}} E(M_{[e_1,\ ...\ , e_k]})$$
Let $L_{n}(M)$ be $\neg{\left\langle -\top\right\rangle}^{n}\top$, $L(M)$ be $E(M)\wedge\bigwedge_{1\leq i\leq n}L_i(M)$. Let the symbol $\underline{\leftrightarrow}_r$ indicate the existence of a r-bisimulation between two pointed models.
\begin{theorem}
	For any two pointed models $(M,s)=(W_1,R_1,V_1,s)$ and $(N,t)= (W_2,R_2,V_2,t)$, the following are equivalent:
	\begin{itemize}
		\item[$(a)$] $(M,s)\underline{\leftrightarrow}_r (N,t)$
		\item[$(b)$] $(N,t)$ can be expanded to a model $(N^\prime,t)= (W_2,R_2,V_{2}^{\prime},t)$ for $L(M)$ such that $p_s$ holds at world $t$.
	\end{itemize} 
\end{theorem}
\begin{proof}
	$[(a)\Longrightarrow(b)]$ Define $N^\prime=(W_2,R_2,V^\prime_2)$, where $V_{2}^{\prime}(p)= V_2 (p)$ for any proposition letter in the initial language,   $V_{2}^{\prime}(p_x)=\{u|(M,x)\underline{\leftrightarrow}_r (N,u) \}$ for $x\in W_1$. We have to prove  
	$N^\prime\vDash L(M)\And N^\prime, t\vDash p_{s}$. Since  $(M,s)\underline{\leftrightarrow}_r (N,t)$, then $t\in V_{2}^{\prime}(p_s)$, which means $p_s$ holds at world $t$. We have to prove $N^\prime\vDash L(M)$. Since $L(M)$ is $E(M)\wedge\bigwedge_{1\leq i\leq n}L_i(M)$. Firstly, we prove that  $N^\prime\nvDash E(M)$ by contradiction.
	
	Since $E(M)$ is the conjuction of all statements $p_x\rightarrow AT_x\wedge env(M,x)$, where $env(M,x)$ is the formula $\bigwedge_{y\in \{y|R_1 xy\}}\Diamond p_{y} \wedge \Box\bigvee_{y\in \{y|R_1 xy\}} p_{y}$.
	Suppose $N^\prime\nvDash E(M)$, then there exists $e$ in $N^\prime$ such that $N^\prime, e \nvDash p_y\rightarrow AT_y \wedge env(M,y)$ for some $y\in W_1$, then $N^\prime, e \vDash p_y$ and $N^\prime, e \nvDash AT_y \wedge env(M,y)$. Since $N^\prime, e \vDash p_y$, then $(M,y)\underline{\leftrightarrow}_r (N,e)$, then $(M,y)\vDash p$ iff $N,e\nvDash p$ for any proposition letter $p$ in the initial language.
	\begin{itemize}
		\item [$\bullet$] If $N^\prime, e \nvDash AT_y$, then there exists $p$ in the initial language such that $(M,y)\vDash p$ iff $N^\prime,e\nvDash p$ iff $N,e\nvDash p$, contradiction.
		\item [$\bullet$] If $N^\prime, e \nvDash  env(M,y)$, then $N^\prime, e \nvDash \bigwedge_{z\in \{y|R_{1} yz\}}\Diamond p_{z} \wedge \Box\bigvee_{z\in \{y|R_{1} yz\}} p_{z}$.
		\begin{itemize}
			\item[$-$] If $N^\prime, e \nvDash \bigwedge_{z\in \{y|R_{1} yz\}}\Diamond p_{z} $, then  $N^\prime, e \nvDash \Diamond p_z$ for some $z$ with $R_{1}yz$, since $(M,y)\underline{\leftrightarrow}_r (N,e)$, then there exists $z^\prime$ with $R_2e{z}^\prime$ and $(M,z)\underline{\leftrightarrow}_r (N,z^\prime)$, then $z^\prime\in V_{2^{\prime}}(p_z)$, thus $N,z^\prime\vDash p_z$, then we have $N,e\vDash\Diamond p_z$, contradiction.
			\item[$-$] If $N^\prime, e \nvDash \Box\bigvee_{z\in \{z|R_{1} yz\}} p_{z}$, then there exists world $e^\prime$ with $R_{2}ee^\prime$, such that $N^\prime, e^\prime \vDash \bigwedge_{z\in \{y|R_{1} yz\}}\neg p_z$. since $(M,y)\underline{\leftrightarrow}_r (N,e)$, then there exists $y^\prime$ with $R_{1}yy^\prime$ such that $(M,y^\prime)\underline{\leftrightarrow}_r (N,e^\prime)$, then $p_{y^\prime}$ holds at world $e^\prime$, contradiction.
		\end{itemize}
	\end{itemize}
	Then we prove $N^\prime\vDash\bigwedge_{1\leq i\leq n}L_i(M)$ directly. For any  point-deleting sequence $(d_1,...,d_l)$ satisfying  $Q_{(d_1,...,d_l)}$ from  $M$, because of the $Zig_{\left\langle -\psi\right\rangle }$ item of r-bisimulation, there exists a point-deleting sequence $(j_1,...,j_l)$ satisfying  $Q_{(j_1,...,j_l)}$ from  $N$, such that for $0\leq m<l$, $(M_{[d_1,...,d_m]},d_{m+1})\underline{\leftrightarrow}_r (N_{[j_1,...,j_m]},j_{m+1})$ $(M_{[d_1,...,d_m]},s)\underline{\leftrightarrow}_r (N_{[j_1,...,j_m]},t)$. Similarly with the first part, we have that $(N_{[j_1,...,j_l]},t)$ can be expanded to a model $({N_{[j_1,...,j_l]}}^\prime,t)$ for $E(M_{[d_1,...,d_l]})$, since ${N_{[j_1,...,j_l]}}^\prime= N_{[j_1,...,j_l]}^\prime$, then $ N^\prime\vDash  {\left\langle -p_{d_1}\right\rangle }...\ {\left\langle -p_{d_l}\right\rangle }E(M_{[(d_1,...,d_l)]})$, then $ N^\prime\vDash \bigwedge_{P_{(i_1,...,i_l)}} {\left\langle -p_{i_1}\right\rangle }...\ {\left\langle -p_{i_l}\right\rangle }$
	
	\noindent $E(M_{[i_1,\ ...\ , i_l]})$. Since for any point-deleting sequence $(j_1,...,j_l)$ satisfying  $Q_{(j_1,...,j_l)}$ from  $N$, thus $ N_{[j_1,...,j_l]}^\prime\vDash \bigvee_{P_{(i_1,...,i_l)}} E(M_{[i_1,\ ...\ , i_l]})$, then $ N^\prime\vDash  [-p_{i_1}]...\ [-p_{i_l}]\bigvee_{P_{(i_1,...,i_l)}} E(M_{[i_1,\ ...\ , i_l]})$. And since model $N$ has the same number of edges with model $M$, Thus $N^\prime\vDash \neg{\left\langle -\top\right\rangle}^{n}\top$, then $N^\prime\vDash L(M)$.
	
	\smallskip
	
	$[(b)\Longrightarrow(a)]$It is sufficient to prove Lemma $18$ and $19$ given below.
\end{proof}	

\begin{lemma}
	For any point-deleting sequence $(d_1,...,d_l)$ satisfying $Q_{(d_1,...,d_l)}$ from  $M$, where $0\leq l< n$, if there is a path $s_0\rightarrow s_1\rightarrow\ ...\ \rightarrow s_k$ in $M_{[d_1,...,d_l]}$, then there exists a point-deleting sequence $(j_1,...,j_l)$ from  $N$ satisfying $p_{d_m}$ holds at $j_m$  for $1 \leq m\leq l$, and a path  $t_0\rightarrow t_1\rightarrow\ ...\ \rightarrow t_k$ in $N_{[j_1,...,j_l]}$   such that $s_i\in V_{1}(p)$ iff $t_i\in V_{2}(p)$ for each proposition letter $p$ in the initial language, where $s_0$ is reachable from $s$ in $M$ , $t_0$ is reachable from $t$ in $N$, $i\in [0,k]$. 		
\end{lemma}

\begin{proof}
	For any path $s_0\rightarrow s_1\rightarrow\ ...\ \rightarrow s_k$ in $M_{[d_1,...,d_l]}$, since $ N^\prime\vDash \bigwedge_{Q_{(i_1,...,i_l)}} {\left\langle -p_{i_1} \right\rangle }... {\left\langle -p_{i_l} \right\rangle }E(M_{[i_1,\ ...\ , i_l]})$, then $ N^\prime\vDash {\left\langle -p_{d_1} \right\rangle }... {\left\langle -p_{d_l} \right\rangle }E(M_{[d_1,...,d_l]})$, thus  $ N^\prime_{[j_1,...,j_l]}\vDash E(M_{[d_1,...,d_l]})$ for some point-deleting sequence $(j_1,...,j_l)$ from  $N^\prime$ satisfing $p_{d_m}$ holds at $j_m$  for $1 \leq m\leq l$.	We take $s_0$ satisfying $N^\prime,t_0\vDash p_{s_0}$, since $N^\prime_{[j_1,...,j_l]}, t_0\vDash p_{s_0}\rightarrow AT_{s_0} \wedge env(M_{[d_1,...,d_l]},s)$, then $N^\prime,t_0\vDash AT_{s_0}$. Suppose that we have select $t_i(i<k)$ satisfying $N_{[j_1,...,j_l]}^\prime,t_i\vDash  p_{s_i}$,  we prove that there exists $t_{i+1}$ such that $N^\prime_{[j_1,...,j_l]}, t_{i+1}\vDash p_{s_{i+1}}$ and $R_{2}t_{i}t_{i+1}$. Since $N^\prime_{[j_1,...,j_l]}, t_i\vDash p_{s_i}\rightarrow AT_{s_i} \wedge env(M_{[d_1,...,d_l]},s_i)$, then $N^\prime_{[j_1,...,j_l]},t_i\vDash AT_{s_i}$, and $N^\prime_{[j_1,...,j_l]},t_i\vDash env(M_{[d_1,...,d_l]},s_i)$. Thus we have $N^\prime_{[j_1,...,j_l]},t_i\vDash \Diamond p_{s_{i+1}}$, then  there exists $e$ with $R_{2} t_{i}e$ such that $N^\prime_{[j_1,...,j_l]},e\vDash p_{s_{i+1}}$, since 	$N^\prime_{[j_1,...,j_l]}, e\vDash p_{s_{i+1}}\rightarrow AT_{s_{i+1}} \wedge env(M_{[d_1,...,d_l]},$
	
	\noindent $s_{i+1})$, then $N^\prime_{[j_1,...,j_l]}, e\vDash AT_{s_{i+1}}$. Let $t_{i+1}$ be $e$, then we finish our lemma.
\end{proof}

\begin{lemma}
	For any point-deleting sequence $(j_1,...,j_l)$ satisfying $Q_{(j_1,...,j_l)}$ from  $N$, where $0\leq l< n$, if there is a path $t_0\rightarrow t_1\rightarrow\ ...\ \rightarrow t_k$ in $N_{[j_1,...,j_l]}$ , then there exists a point-deleting sequence $(d_1,...,d_l)$ from  $M$ satisfying $p_{d_m}$ holds at $j_m$  for $1 \leq m\leq l$ and a path $s_0\rightarrow s_1\rightarrow\ ...\ \rightarrow s_k$ in $M_{[d_1,...,d_l]}$  such that $s_i\in V_{1}(p)$ iff $t_i\in V_{2}(p)$ for each proposition letter $p$ in the initial language, where $t_0$ is reachable from $t$ in $N$, $s_0$ is reachable from $s$ in $M$, $i\in [0,k]$.
\end{lemma}

\begin{proof}
	Since $N^\prime\vDash\bigwedge_{Q_{(e_1,...,e_n)}} {\left\langle -p_{e_1}\right\rangle }...\ {\left\langle -p_{e_n}\right\rangle }E(M_{[e_1,\ ...\ , e_n]})$, then for any world $v$ in $N^\prime$, it must be statisfied by $p_w$ for some world $v$ in $M$. Since 
	 $ N^\prime\vDash {[-i_1]...[-i_l]}\bigvee_{P_{(i_1,...,i_l)}} E(M_{[i_1,\ ...\ , i_l]})$, then for any point-deleting sequence $(j_1,...,j_l)$ from  $N^\prime$,  $N_{[j_1,...,j_l]}^\prime \vDash E(M_{[d_1,...,d_l]})$ for some point-deleting sequence $(d_1,...,d_l)$ from  $M$ satisfying $p_{d_m}$ holds at $j_m$  for $1 \leq m\leq l$. For any path $t_0\rightarrow t_1\rightarrow\ ...\ \rightarrow t_k$ in $N_{[j_1,...,j_l]}$, we select $s_0$ satisfing $N_{[j_1,...,j_l]}^\prime,t_0\vDash p_{s_{0}}$, since 
	$N^\prime_{[j_1,...,j_l]}, t_0\vDash p_{s_0}\rightarrow AT_{s_0} \wedge env(M_{[d_1,...,d_l]},s_0)$, then $N^\prime_{[j_1,...,j_l]},t_0\vDash AT_{s_0}$.  Suppose that we have select $s_i(i<k)$ satisfying $N_{[j_1,...,j_l]}^\prime,t_i\vDash  p_{s_i}$, we prove that there exists $s_{i+1}$ such that $N^\prime, t_{i+1}\vDash p_{s_{i+1}}$ and $R_{1}s_{i}s_{i+1}$. Since $N^\prime_{[j_1,...,j_l]}, t_i\vDash p_{s_i}\rightarrow AT_{s_i} \wedge env(M_{[d_1,...,d_l]},s_i)$, then $N^\prime_{[j_1,...,j_l]},t_i\vDash AT_{s_i}$, and $N^\prime_{[j_1,...,j_l]},t_i\vDash env(M_{[d_1,...,d_l]},s_i)$. If $env(M_{[d_1,...,d_l]},s_i)$ is $\top\wedge\Box\bot$(when $s_i$ has no successor in $M_{[d_1,...,d_l]}$),  then $N^\prime_{[j_1,...,j_l]},t_i\vDash\Box\bot$, it's impossible since $t_{i+1}$ is a successor of world $t_i$ in $N^\prime_{[j_1,...,j_l]}$, then there exists $z$ with $R_{1}s_{i}z$ such that $N^\prime_{[j_1,...,j_l]}, t_{i+1}\vDash p_{z}$, since 
	$N^\prime_{[j_1,...,j_l]}, t_{i+1}\vDash p_{z}\rightarrow AT_{z} \wedge env(M_{[d_1,...,d_l]},z)$, then $N^\prime_{[j_1,...,j_l]}, t_{i+1}\vDash  AT_{z}$. Let $s_{i+1}$ be  $z$, we finish our proof.
\end{proof}


\newpage
\section*{Appendix B: Algorithms for finding bisimulations}

    \subsection*{Correctness of Algorithm 1}
    The following lemma will be useful in the correctness proof.
    \begin{lemma}
        If $((W_1,R_1,V_1),w_1)\underline{\leftrightarrow}_s((W_2,R_2,V_2),w_2)$ and $|R_1|$ and $|R_2|$ are finite, then $|R_1|=|R_2|$.
    \end{lemma}
    \begin{proof}
        Suppose on the contrary, $|R_1|\neq |R_2|$. Without loss of generality, assume $|R_1|<|R_2|$.\\
        Proof by induction on $n=|R_1|$
        \begin{itemize}
            \item \textbf{Base case:} $n=0$\\
                By assumption $|R_1|=0$ and $|R_2|>0$. So $\exists e\in R_2$. Now, since $((W_1,R_1,V_1),w_1)\underline{\leftrightarrow}_s((W_2,R_2,V_2),w_2)$, they satisfy condition (5) of the definition of s-bisimilarity. Therefore, there must exist an edge $f\in R_1$ such that $((W_1,R_1\backslash\{f\},V_1),w_1)\underline{\leftrightarrow}_s((W_2,R_2\{e\},V_2),w_2)$. But, since $|R_1|=0$, no such $f$ can exist. contradiction.
            \item \textbf{Induction hypothesis:} Suppose the claim holds good for $n\leq k$, $i.e.,$ $|R_1|=|R_2|$, whenever $|R_1|\leq k$
            \item \textbf{Induction step:} $n=k+1$\\
                Suppose $\min(|R_1|, |R_2|)=|R_1|=k+1$. Let $e_1\in R_1$ be any edge. Since, $((W_1,R_1,V_1),w_1)\underline{\leftrightarrow}_s((W_2,R_2,V_2),w_2)$, they satisfy condition (4) in the definition of s-bisimilarity, so there exists $e_2\in R_2$ such that $((W_1,R_1\backslash\{e_1\},V_1),w_1)$ $\underline{\leftrightarrow}_s((W_2,R_2\backslash\{e_2\},V_2),w_2)$. But then by induction hypothesis, we have $|R_1\backslash\{e_1\}|=R_2\backslash\{e_2\}|\implies|R_1|-1=|R_2|-1\implies |R_1|=|R_2|$ 
        \end{itemize}
        This completes the proof.
    \end{proof}
    
    \noindent Now we give the correctness proof.
        \begin{theorem}
               Given two models $(\mathcal{M}_1, w_1)$ and $(\mathcal{M}_2, w_2)$, where $\mathcal{M}_1=(W_1,R_1,V_1)$, $\mathcal{M}_2=(W_2,R_2,V_2)$, $w_1\in W_1$ and $w_2\in W_2$; $(\mathcal{M}_1, w_1)\underline{\leftrightarrow}_s (\mathcal{M}_2, w_2)$ iff the function 
               \textsf{s-bisimilar}($(\mathcal{M}_1, w_1), (\mathcal{M}_2, w_2), \emptyset)$ returns yes.
        \end{theorem}
        \begin{proof}
            Suppose $\mathcal{M}_1$ and $\mathcal{M}_2$ have different number of edges, then s-bisimilar(($(\mathcal{M}_1, w_1), (\mathcal{M}_2, w_2), \emptyset)$ returns No at line 4, and $(\mathcal{M}_1, w_1)\cancel{$\underline{\leftrightarrow}$}_s (\mathcal{M}_2, w_2)$.
       So, let us consider that both models have equal number of edges (say n).
           We prove by induction on n:
            \begin{itemize}
                \item[>] \textbf{Base case:} $n=0$.\\
                To prove $(\mathcal{M}_1, w_1)\underline{\leftrightarrow}_s (\mathcal{M}_2, w_2)$ iff the function \textsf{s-bisimilar}($(\mathcal{M}_1, w_1), (\mathcal{M}_2, w_2), \emptyset)$ returns yes when $R_1=\emptyset=R_2$. We will first prove, by contrapositivity, that if \textsf{s-bisimilar}($(\mathcal{M}_1, w_1), (\mathcal{M}_2, w_2), \emptyset)$ returns yes, then $(\mathcal{M}_1, w_1)\underline{\leftrightarrow}_s (\mathcal{M}_2, w_2)$.
                    \item[> >] Suppose $(\mathcal{M}_1, w_1)\cancel{$\underline{\leftrightarrow}$}_s (\mathcal{M}_2, w_2)$. Then they violate one of the five conditions in the definition of s-bisimilarity (in section 2.1.3).
                        \item[> > >] Suppose they violate condition (1). There there is some atomic proposition $p$ such that either $(\mathcal{M}_1,w_1)\models p$ and $(\mathcal{M}_2,w_2)\not\models p$; or $(\mathcal{M}_1,w_1)\not \models p$ and $(\mathcal{M}_2,w_2)\models p$. From truth definition of SML, we have $w_1\in V_1(p)$ but $w_2\not\in V_2(p)$; or $w_1\not\in V_1(p)$ but $w_2\in V_2(p)$. In this case the function returns NO in line 7.
                        \item[> > >] Suppose they violate condition (2). Then, there is a successor $v_1$ of $w_1$, i.e. $\exists v_1\in W_1$ such that $w_1R_1v_1$, but $\forall v_2$ such that $w_2R_2v_2$, we do not have $(\mathcal{M}_1, v_1)\underline{\leftrightarrow}_s (\mathcal{M}_2, v_2)$. But since $n=0$, $w_1R_1v_1$ does not hold for any $v_1$ as $R_1=\emptyset$. Therefore, condition (2) in the definition of s-Bisimilarity cannot be violated in this case.
                        \item[> > >] Suppose that they violate condition (3). Again by similar argument as last point, we get can not have $v_2R_2w_2$ and hence condition (3) can not be violated when $n=0$.
                        \item[> > >] Suppose they violate condition (4). Then there is an edge $e_1\in R_1$ such that for any edge $e_2\in R_2$, it is not the case that $(\mathcal{M}_1\backslash\{e_1\}, w_1)\underline{\leftrightarrow}_s (\mathcal{M}_2\backslash\{e_2\}, w_2)$. But again since $n=0$, $R_1 = \emptyset$, hence no such $e_1$ exists. So this case cannot arise.
                        \item[> > >] By similar argument as in previous point, the models cannot violate condition (5).
                    Now we will prove the other side, again by contrapositivity.
                    \item[> >] Conversely, suppose that s-bisimilar(($(\mathcal{M}_1, w_1), (\mathcal{M}_2, w_2), \emptyset)$ returns No, Then one of the following cases occur:
                        \item[> > >] The function returns No at line number 7. This can only happen when the If condition in line 6 is true. Therefore, there exists an atomic proposition $p$ such that, $w_1\in V_1(p)$ but $w_2\not\in V_2(p)$; or $w_1\not\in V_1(p)$ but $w_2\in V_2(p)$. From truth definition of SML, we have either $(\mathcal{M}_1,w_1)\models p$ and $(\mathcal{M}_2,w_2)\not\models p$; or $(\mathcal{M}_1,w_1)\not \models p$ and $(\mathcal{M}_2,w_2)\models p$. But then $(\mathcal{M}_1, w_1)\cancel{$\underline{\leftrightarrow}$}_s (\mathcal{M}_2, w_2)$ as they violate condition (1) of the definition of s-bisimilarity.
                        \item[> > >] The function returns No at line number 15. But since $R_1=\emptyset$, the body of for loop at line 8 is not executed. Hence line 15 is not executed.
                        \item[> > >] By similar argument as previous case, line 23 is not executed and hence function cannot return No from line 23
                        \item[> > >] Suppose the function returns no from line 35. then the condition at line 34 is true. Therefore $(w_1R_1u_1)$ is true. But this cannot be the case as $R_1=\emptyset$.
                        \item[> > >] By similar argument as previous case, the function cannot return No from line 46
                This completes both sides of the base case.
                \item[>] \textbf{Induction Hypothesis:} Suppose the theorem holds good for $n\leq k$. That is, $(\mathcal{M}_1, w_1)\underline{\leftrightarrow}_s (\mathcal{M}_2, w_2)$ iff the function \textsf{s-Bisimilar}($(\mathcal{M}_1, w_1), (\mathcal{M}_2, w_2), \emptyset)$ returns yes when $|R_1|=|R_2|\leq k$
                \item[>] \textbf{Induction Step:} Let $n=k+1$\\
                We will first prove that if $(\mathcal{M}_1, w_1)\underline{\leftrightarrow}_s (\mathcal{M}_2, w_2)$ then the function \textsf{s-Bisimilar}($(\mathcal{M}_1, w_1), (\mathcal{M}_2, w_2), \emptyset)$ returns yes. Again we will prove this by contrapositivity.
                    \item[> >] Suppose the function returns No. Then it executes one of the 6 return No statements. But it can not return NO at line 4, as we have assumed $|R_1|=|R_2|$. So the following cases can occur:
                        \item[> > >] The function returns No at line number 7. This can only happen when the If condition in line 6 is true. But then, by argument similar to that in base case, $(\mathcal{M}_1, w_1)\cancel{$\underline{\leftrightarrow}$}_s (\mathcal{M}_2, w_2)$ as they violate condition (1) of the definition of s-bisimilarity.
                        \item[> > >] The function returns No at line number 15. Then condition in line 14 is true even after execution of forloop at line 10. Therefore, there is some $e_1\in R_1$ such that for all $e_2\in R_2$, condition in line 11 is false, $i.e.$ there is an $e_1\in R_1$ such that for all $e_2\in R_2$, we have \textsf{s-Bisimilar}($(\mathcal{M}_1\backslash\{e_1\},w_1), (\mathcal{M}_2\backslash\{e_2\},w_2), \emptyset)$ returns NO. But the model $\mathcal{M}_1^\prime=\mathcal{M}_1\backslash\{e_1\}$ and $\mathcal{M}_2^\prime=\mathcal{M}_2\backslash\{e_2\}$ have k edges. Therefore, by induction hypothesis, $(\mathcal{M}_1\backslash\{e_1\}, w_1)\cancel{$\underline{\leftrightarrow}$}_s (\mathcal{M}_2\backslash\{e_2\}, w_2)$ for all $e_2\in R_2$. This is violation to condition (4) in the definition of s-bisimilarity. Therefore, $(\mathcal{M}_1, w_1)\cancel{$\underline{\leftrightarrow}$}_s (\mathcal{M}_2, w_2)$
                        \item[> > >] The function returns no on line 23. By similar argument as in previous case, this leads to violation of condition (5) in the definition of s-bisimilarity. Hence, $(\mathcal{M}_1, w_1)\cancel{$\underline{\leftrightarrow}$}_s (\mathcal{M}_2, w_2)$
                        \item[> > >] The function returns No at line 35. Then condition at line 34 is true for some $u_1\in W_1$. Therefore, following cases arise:
                            \item[> > > >] For a successor $u_1$ of $w_1$, condition at line 28 is false for all $u_2\in W_2$, i.e., $w_2R_2u_2$ is not true for any $u_2\in W_2$. This is a violation of condition (2) in definition of s-bisimilarity and hence $(\mathcal{M}_1, w_1)\cancel{$\underline{ \leftrightarrow}$}_s (\mathcal{M}_2, w_2)$
                            \item[> > > >] Condition at line 29 is true but condition at line 30 is false, i.e., $\exists u_1\in W_1$ such that $w_1R_1u_1$, $\forall u_2\in R_2$ such that $w_2R_2u_2$ and $L$ is such that $(u_1, u_2)\not\in L$ (and $(w_1,w_2)\not\in L$ because line 29 can be executed only if condition in line 24 is true); we get \textsf{s-Bisimilar}($(\mathcal{M}_1, u_1), (\mathcal{M}_2, u_2), L\cup\{(w_1,w_2)\})$) returns No.\\
                            To prove:$(\mathcal{M}_1, w_1)\cancel{$\underline{\leftrightarrow}$}_s (\mathcal{M}_2, w_2)$.\\
                            Proof by induction on $m=|W_1\times W_2|-|L\cup\{(w_1,w_2)\}|$
                                \item[> > > > >] \textbf{Base case:} $|W_1\times W_2|=|L\cup\{(w_1,w_2)\}|$\\
                                We need to prove that if $\exists u_1\in W_1$ such that $w_1R_1u_1$, $\forall u_2\in R_2$ such that $w_2R_2u_2$ and $L$ is such that $(u_1, u_2)\not\in L$ (and $(w_1,w_2)\not\in L$ because line 29 can be executed only if condition in line 24 is true) and $|W_1\times W_2|-|L\cup\{(w_1,w_2)\}|=0$; and \textsf{s-Bisimilar}($(\mathcal{M}_1, u_1), (\mathcal{M}_2, u_2), L\cup\{(w_1,w_2)\})$) returns No, then $(\mathcal{M}_1, w_1)\cancel{$\underline{\leftrightarrow}$}_s (\mathcal{M}_2, w_2)$\\
                                But since $|W_1\times W_2|-|L\cup\{(w_1,w_2)\}|=0$, we have $(u_1,u_2)\in L\cup\{(w_1,w_2)\}$. This is in contradiction with condition in line 29 being true. So the antecedent is false and hence base case is true vacuously.
                                \item[> > > > >] \textbf{Induction Hypothesis:} Suppose the claim holds for $m\leq l$, $i.e.$, \\
                                Suppose whenever $\exists u_1\in W_1$ such that $w_1R_1u_1$, $\forall u_2\in R_2$ such that $w_2R_2u_2$ and $L$ is such that $(u_1, u_2)\not\in L$ (and $(w_1,w_2)\not\in L$ because line 29 can be executed only if condition in line 24 is true) and $|W_1\times W_2|-|L\cup\{(w_1,w_2)\}|\leq l$; and \textsf{s-Bisimilar}($(\mathcal{M}_1, u_1), (\mathcal{M}_2, u_2), L\cup\{(w_1,w_2)\})$) returns No, then $(\mathcal{M}_1, w_1) \cancel{$\underline{\leftrightarrow}$}_s (\mathcal{M}_2, w_2)$
                                \item[> > > > >] \textbf{Induction step:} Suppose $m=l+1$.\\
                                In this case, suppose condition in line 29 true and condition in line 30 is false. Therefore, we have, $\exists u_1\in W_1$ such that $w_1R_1u_1$, $\forall u_2\in R_2$ such that $w_2R_2u_2$ and $L$ is such that $(u_1, u_2)\not\in L$ (and $(w_1,w_2)\not\in L$ because line 29 can be executed only if condition in line 24 is true) and $|W_1\times W_2|-|L\cup\{(w_1,w_2)\}|= l+1$; and \textsf{s-Bisimilar}($(\mathcal{M}_1, u_1), (\mathcal{M}_2, u_2), L\cup\{(w_1,w_2)\})$) returns No. Now,  \textsf{s-Bisimilar}($(\mathcal{M}_1, u_1), (\mathcal{M}_2, u_2), L\cup\{(w_1,w_2)\})$) can return No either at one of 6 return No statements. If it returns No at lines 4, 7, 15 or 23, then by above cases, we have already proved that $(\mathcal{M}_1, u_1) \cancel{$\underline{\leftrightarrow}$}_s (\mathcal{M}_2, u_2)$ because they violate conditions (1) or (4) or (5) in the definition of s-bisimilarity. Suppose it returns No at line 35, then if condition at line 28 is always false, then $(\mathcal{M}_1, u_1) \cancel{$\underline{\leftrightarrow}$}_s (\mathcal{M}_2, u_2)$ because they violate condition (2) of definition of s-bisimilarity. So suppose condition at line 29 is true but at line 30 is false. Therefore, $\exists v_1\in W_1$ such that $u_1R_1v_1$ and $\forall v_2\in W_2$ such that $u_2R_2v_2$, $L$ is such that $(v_1,v_2)\not\in L\cup\{(w_1,w_2)\}$ (also $(u_1,u_2)\not\in L\cup\{(w_1,w_2)\}$ because condition at line24 has to be true), we have \textsf{s-Bisimilar}($(\mathcal{M}_1,v_1), (\mathcal{M}_2,v_2), L\cup\{(w_1,w_2), (u_1,u_2)\}$) returns No. Now by induction hypothesis, $(\mathcal{M}_1, v_1) \cancel{$\underline{\leftrightarrow}$}_s (\mathcal{M}_2, v_2)$ which implies $(\mathcal{M}_1, u_1) \cancel{$\underline{\leftrightarrow}$}_s (\mathcal{M}_2, u_2)$ and hence $(\mathcal{M}_1, w_1) \cancel{$\underline{\leftrightarrow}$}_s (\mathcal{M}_2, w_2)$.
                        \item[> > >] The function returns No at line 46, then by argument similar to last case, $(\mathcal{M}_1, w_1) \cancel{$\underline{\leftrightarrow}$}_s (\mathcal{M}_2, w_2)$.
                    We will now prove the remaining side by contrapositivity.
                    \item[>] Suppose $(\mathcal{M}_1, w_1)\cancel{$\underline{\leftrightarrow}$}_s (\mathcal{M}_2, w_2)$. Then these models must violate one of the 5 conditions in definition of s-bisimilarity.
                        \item[> >] Suppose they violate condition (1). There there is some atomic proposition $p$ such that either $(\mathcal{M}_1,w_1)\models p$ and $(\mathcal{M}_2,w_2)\not\models p$; or $(\mathcal{M}_1,w_1)\not \models p$ and $(\mathcal{M}_2,w_2)\models p$. From truth definition of SML, we have $w_1\in V_1(p)$ but $w_2\not\in V_2(p)$; or $w_1\not\in V_1(p)$ but $w_2\in V_2(p)$. In this case the function returns NO in line 7.
                        \item[> >] Suppose they violate condition (4). Then there is an edge $e_1\in R_1$ such that for any edge $e_2\in R_2$, $(\mathcal{M}_1\backslash\{e_1\}, w_1)\cancel{$\underline{\leftrightarrow}$}_s (\mathcal{M}_2\backslash\{e_2\}, w_2)$. In this case for $e_1$, condition in line 11 is never true by induction hypothesis ($\mathcal{M}_1\backslash\{e_1\}$ and $\mathcal{M}_2\backslash\{e_2\}$ have k edges, hence we can use induction hypothesis). Therefore, return No is executed in line 15.
                        \item[> >] Suppose they violate condition (5), by similar argument as previous case, by induction hypothesis, function returns No.
                        \item[> >] Suppose they violate condition (2) and/or (3). We need to prove if $(\mathcal{M}_1,w_1)\cancel{$\underline{\leftrightarrow}$}_s(\mathcal{M}_2,w_2)$ because they violate condition (2) and/or, but not (1), (4) or (5) in the definition of s-bisimilarity, then
                        s-bisimilar(($(\mathcal{M}_1, w_1)$ ,$ (\mathcal{M}_2, w_2), L)$ returns No, for $|L|=0$.\\
                        Since $(\mathcal{M}_1,w_1)\cancel{$\underline{\leftrightarrow}$}_s(\mathcal{M}_2,w_2)$ because they violate condition (2) and/or (3), therefore $\exists u_{11}\in W_1$, $w_1R_1u_{11}$, such that $\forall u_{21}\in W_2$, $w_2R_2u_{21}$, $(\mathcal{M}_1, u_{11})\cancel{$\underline{\leftrightarrow}$}_s (\mathcal{M}_2, u_{21})$ (if condition (2) is violated); or  $\exists u_{12}\in W_2$, $w_2R_2u_{12}$, such that $\forall u_{11}\in W_1$, $w_1R_1u_{11}$, $(\mathcal{M}_1, u_{11})\cancel{$\underline{\leftrightarrow}$}_s (\mathcal{M}_2, u_{21})$. Now if $(\mathcal{M}_1, u_{11})\cancel{$\underline{\leftrightarrow}$}_s (\mathcal{M}_2, u_{21})$ because they violate conditions (1), (4) or (5), then by previous cases, the function returns No at line 7, 15 or 23 respectively and we will be done. Let us pick a general such pair $(v_{11},v{21})$. So, assume $(\mathcal{M}_1, v_{11})\cancel{$\underline{\leftrightarrow}$}_s (\mathcal{M}_2, v_{21})$ because they violate condition(s) (2) and/or (3). Therefore, again, $\exists u_{12}\in W_1$, $v_{11}R_1u_{12}$, such that $\forall u_{22}\in W_2$, $v_{12}R_2u_{22}$, $(\mathcal{M}_1, u_{12})\cancel{$\underline{\leftrightarrow}$}_s (\mathcal{M}_2, u_{22})$ (if they violate (2)); or $\exists u_{22}\in W_2$, $v_{12}R_2u_{22}$, such that $\forall u_{12}\in W_1$, $v_{11}R_1u_{12}$, $(\mathcal{M}_1, u_{12})\cancel{$\underline{\leftrightarrow}$}_s (\mathcal{M}_2, u_{22})$ (if they violate condition (3). Again, choose a general such pair $(v_{12}, v_{22})$ from above such that $v_{11}R_1v_{12}$ and $v_{21}R_2v_{22}$ and $(\mathcal{M}_1, v_{12})\cancel{$\underline{\leftrightarrow}$}_s (\mathcal{M}_2, v_{22})$. Again, we are done if $(\mathcal{M}_1, v_{12})\cancel{$\underline{\leftrightarrow}$}_s (\mathcal{M}_2, v_{22})$ because they violate condition (1), (4) or (5). So, again, we can assume that they violate condition (2) and /or (3). This can go on until we reach a leaf node, i.e., there is some $k$ such that exactly one of the following is true: $v_{1k}R_1v_{1k+1}$ for some $v_{1k+1}\in W_1$ or $v_{2k}R_2v_{2k+1}$ for some $v_{2k+1}\in W_2$. Again the function returns No, either at line 15 or 23 respectively in both cases. The only case that remains is when there is no leaf nodes and there is some $k$ such that $v_{1k}=v_{1l}$ or $w_1$ and $v_{2k}=v_{2l}$ or $w_2$ for some $l<k$. In this case, since $v_{1i}$ and $v_{2i}$ were some general node in the reachable part from $w_1$ and $w_2$, such that they do not violate condition (1), (4) or (5) in the definition of s-bisimilarity, we have the following:
                            \item[> > >] $(\mathcal{M}_1,w_1)$ and $(\mathcal{M}_2,w_2)$ satisfy conditions (1), (4) and (5) in the definition of s-bisimilarity
                            \item[> > >] For every n, $\exists v_1\in W_1$ such that $w_1R^n_1v_1$ iff $\exists v_2\in W_2$ such that $w_2R^n_2v_2$; and $(\mathcal{M}_1,v_1)$ and $(\mathcal{M}_2,v_2)$ satisfy condition (1), (4) and (5) from the definition of s-bisimilarity.
                        But these conditions are same as the conditions in definition of s-bisimilarity. Hence, $(\mathcal{M}_1, w_1)\underline{\leftrightarrow}_s (\mathcal{M}_2, w_2)$ and the function does not return No in this this case.

            \end{itemize}
            This completes the proof.
        \end{proof}
    \subsection*{An example}
    Below is an example run of the function \textsf{s-Bisimilar}, with input as the models given in root node, pointed at $w_1$ and $w_2$. Proposition $p$ is true in all the worlds of both the models. The recursion graph shows all the important nodes.

\tikzset{every picture/.style={line width=0.75pt}} 
\begin{figure}[H]
\centering


\end{figure}
    

    \section*{Algorithm for point sabotage bisimulation}
    The resulting model after deleting an edge, just makes change in $R$ and resulting accession relation is same as previous with the edge being deleted. But the resulting model after deleting a point from a model is not as trivial. It not only changes the set of worlds $W$, but also changes the accession relation $R$ and the valuation $V$. So, we first describe the algorithm 2 to compute this resultant model after deleting a point (not same as the $w_1$ where the given model is pointed at). With this, on page 22, we give algorithm 3 to check if two given models are d-bisimilar. The algorithm is very similar to that of the algorithm for s-bisimilarity.
        \begin{algorithm}
            \footnotesize
            \LinesNumbered
                \KwIn{ $((W,R,V),w), u$ } 
                \SetKwFunction{FMain}{successor}
                \SetKwProg{Fn}{Function}{:}{}
                \Fn{\FMain{$((W,R,V),w), u$}}
                {
                    $W^\prime=W\backslash\{u\}$\\
                    $R^\prime=R\backslash(\{(u,v)\in W|v\in R\}\cup\{(v,u)\in R| v\in W\})$\\
                    \ForAll{$p\in\mathcal{P}$}
                    {
                        $V^\prime(p)=V(p)\cap W^\prime$
                    }
                    \Return $((W^\prime,R^\prime,V^\prime), w)$
                }
                \caption{algorithm to compute new relational model after point deletion}
            \end{algorithm}

\begin{algorithm}
    \footnotesize
            \LinesNumbered
                \KwIn{ $((W_1,R_1,V_1),w_1), ((W_2,R_2,V_2),w_2)$ } 
                \textbf{Initialize}:L=$\emptyset$\\
                \SetKwFunction{FMain}{d-Bisimilar}
                \SetKwProg{Fn}{Function}{:}{}
                \Fn{\FMain{$((W_1,R_1,V_1),w_1), ((W_2,R_2,V_2),w_2)$, L}}{
                    {\If{$|W_1|\neq |W_2|$}
                    {\Return NO;}
                    }

                    {
                        \ForAll{atomic propositions p}
                        {
                            \If{$(((w_1\in V_1(p)) AND (w_2\not\in V_2(p)))$ OR $((w_1\not\in V_1(p)) AND (w_2\in V_2(p))))$}
                                {
                                    \Return NO;
                                }
                        }
                        \ForAll{$u_1\in W_1$}
                            {
                                Found=0;\\
                                \ForAll{$u_2\in W_2$}
                                    {
                                        \If{$(u_1\neq w_1) AND (u_2\neq w_2)$}
                                        {
                                        \If {\textsf{d-Bisimilar}$((successor((W_1,R_1,V_1),u_1), w_1)$,$(successor((W_2,R_2,V_2),u_2), w_2), \emptyset)$==YES}
                                            {
                                                Increment Found;\\
                                                break;
                                            }
                                    }
                                    }
                                \If{Found =0 $\wedge (u_1\neq w_1)$}
                                    {
                                        \Return No;
                                    }
                            }
                        
                        \ForAll{$u_2\in W_2$}
                            {
                                Found=0;\\
                                \ForAll{$u_1\in W_1$}
                                    {
                                        \If{$(u_1\neq w_1) AND (u_2\neq w_2)$}
                                        {
                                        \If {\textsf{d-Bisimilar}$((successor((W_1,R_1,V_1),u_1), w_1)$,$(successor((W_2,R_2,V_2),u_2), w_2), \emptyset)$==YES
                                        }
                                            {
                                                Increment Found;\\
                                                break;
                                            }
                                    }
                                    }
                                \If{Found =0 $\wedge (u_2\neq w_2)$}
                                    {
                                        \Return No;
                                    }
                            } \If {$(w_1,w_2)\not\in L$}
                            {
                            \ForAll{$u_1\in W_1$}
                            {
                                Found=0;\\
                                \ForAll{$u_2\in W_2$}
                                    {
                                        \If {($(w_1R_1u_1)$ AND $(w_2R_2u_2))$}
                                        {
                                            \If{$((u_1,u_2)\not\in L)$}
                                            {
                                                \If {\textsf{d-Bisimilar}$(((W_1,R_1,V_1),u_1),((W_2,R_2,V_2),u_2),L\cup\{(w_1,w_2)\})$==YES}
                                                {
                                                    Increment Found;
                                                }
                                            }
                                            \Else{Increment Found}
                                        }
                                            
                                    }
                                \If{(Found=0) AND $(w_1R_1u_1)$}
                                    {
                                        \Return No;
                                    }
                            }
                            
                        \ForAll{$u_2\in W_2$}
                            {
                                Found=0;\\
                                \ForAll{$u_1\in W_1$}
                                    {
                                        \If {($(w_1R_1u_1)$ AND $(w_2R_2u_2))$}
                                        {
                                            \If{$((u_1,u_2)\not\in L)$}
                                            {
                                                \If {\textsf{d-Bisimilar}$(((W_1,R_1,V_1),u_1),((W_2,R_2,V_2),u_2),L\cup\{(w_1,w_2)\})$==YES}
                                                {
                                                    Increment Found;
                                                }
                                            }
                                            \Else{Increment Found}
                                        }
                                            
                                    }
                                \If{(Found=0) AND $(w_2R_2u_2)$}
                                    {
                                        \Return No;
                                    }
                            }
                        }
                        
                        \Return Yes;
                    }
                }
                \caption{Algorithm to check if given two relational modals are d-Bisimilar}
            \end{algorithm}
    
    \begin{lemma}
        If $((W_1,R_1,V_1),w_1)\underline{\leftrightarrow}_d((W_2,R_2,V_2),w_2)$ and $|W_1|$ and $|W_2|$ are finite, then $|W_1|=|W_2|$.
    \end{lemma}
    \begin{proof}
        Suppose on the contrary, $|W_1|\neq |W_2|$. Without loss of generality, assume $|W_1|<|W_2|$.\\
        Proof by induction on $n=|W_1|$
        \begin{itemize}
            \item \textbf{Base case:} $n=1$\\
                By assumption $|W_1|=1$ and $|W_2|>1$. So $\exists u_2\in W_2$ and $u_2\not=w_2$. Now, since $((W_1,R_1,V_1),w_1)\underline{\leftrightarrow}_d((W_2,R_2,V_2),w_2)$, they satisfy condition (5) of the definition of d-bisimilarity. Therefore, there must exist a point $u_1\in W_1$ such that $u_1\neq w_1$. But, since $|W_1|=1$, no such $u_1$ can exist. contradiction.
            \item \textbf{Induction hypothesis:} Suppose the claim holds good for $n\leq k$, $i.e.,$ $|W_1|=|W_2|$, whenever $|W_1|\leq k$
            \item \textbf{Induction step:} $n=k+1$\\
                Suppose $\min(|W_1|, |W_2|)=|W_1|=k+1$. Let $u_1\in W_1$ be any point such that $u_1\neq w_1$. Since, $((W_1,R_1,V_1),w_1)$ $\underline{\leftrightarrow}_d((W_2,R_2,V_2),w_2)$, they satisfy condition (4) in the definition of d-bisimilarity, so there exists $u_2\in W_2$ and $u_2\neq w_2$ such that $(\mathcal{M}_1^\prime,w_1)$ $\underline{\leftrightarrow}_d(\mathcal{M}_2^\prime,w_2)$, where $\mathcal{M}_1^\prime$ and $\mathcal{M}_2^\prime$ are resulting models after deleting points $u_1$ and $u_2$ from $(W_1,R_1,V_1)$ and $(W_2,R_2,V_2)$ respectively. But then by induction hypothesis, we have $|W_1\backslash\{u_1\}|=W_2\backslash\{u_2\}|\implies|W_1|-1=|W_2|-1\implies |W_1|=|W_2|$
        \end{itemize}
        This completes the proof.
    \end{proof}

                \begin{theorem}
               Given two models $(\mathcal{M}_1, w_1)$ and $(\mathcal{M}_2, w_2)$, $(\mathcal{M}_1, w_1)\underline{\leftrightarrow}_p (\mathcal{M}_2, w_2)$ iff the function 
               \textsf{d-Bisimilar}($(\mathcal{M}_1, w_1),$
               
               \noindent $ (\mathcal{M}_2, w_2), \emptyset)$ returns yes.
        \end{theorem}
        \begin{proof}
            The proof is very similar to the correctness of Algorithm 1.\\
            Suppose $\mathcal{M}_1$ and $\mathcal{M}_2$ have different number of points, then d-bisimilar(($(\mathcal{M}_1, w_1), (\mathcal{M}_2, w_2), \emptyset)$ returns No at line 4, and $(\mathcal{M}_1, w_1)\cancel{$\underline{\leftrightarrow}$}_d (\mathcal{M}_2, w_2)$.
       So, let us consider that both models have equal number of points (say n).
           We prove by induction on n:
            \begin{itemize}
                \item[>] \textbf{Base case:} $n=1$.\\
                To prove $(\mathcal{M}_1, w_1)\underline{\leftrightarrow}_d (\mathcal{M}_2, w_2)$ iff the function \textsf{d-bisimilar}($(\mathcal{M}_1, w_1), (\mathcal{M}_2, w_2), \emptyset)$ returns yes when $|W_1|=|W_2|=1$. We will first prove, by contrapositivity, that if \textsf{d-bisimilar}($(\mathcal{M}_1, w_1), (\mathcal{M}_2, w_2), \emptyset)$ returns yes, then $(\mathcal{M}_1, w_1)\underline{\leftrightarrow}_d (\mathcal{M}_2, w_2)$.
                    \item[> >] Suppose $(\mathcal{M}_1, w_1)\cancel{$\underline{\leftrightarrow}$}_d (\mathcal{M}_2, w_2)$. Then they violate one of the five conditions in the definition of d-bisimilarity (in section 2.2.3).
                        \item[> > >] Suppose they violate condition (1). Then there is some atomic proposition $p$ such that either $(\mathcal{M}_1,w_1)\models p$ and $(\mathcal{M}_2,w_2)\not\models p$; or $(\mathcal{M}_1,w_1)\not \models p$ and $(\mathcal{M}_2,w_2)\models p$. From truth definition of Deletion modal logic, we have $w_1\in V_1(p)$ but $w_2\not\in V_2(p)$; or $w_1\not\in V_1(p)$ but $w_2\in V_2(p)$. In this case the function returns NO in line 7.
                        \item[> > >] Suppose they violate condition (2). Then, there is a successor $v_1$ of $w_1$, i.e. $\exists v_1\in W_1$ such that $w_1R_1v_1$, but $\forall v_2$ such that $w_2R_2v_2$, we do not have $(\mathcal{M}_1, v_1)\underline{\leftrightarrow}_d (\mathcal{M}_2, v_2)$. But since $n=1$, $w_1R_1v_1\implies v_1=w_1$ as $|W_1|=1$ and $w_2R_2v_2\implies v_2=w_2$ as $|W_2|=1$. Therefore, $(\mathcal{M}_1, v_1)\underline{\leftrightarrow}_d (\mathcal{M}_2, v_2)$ holds by assumption which is a contradiction.
                        \item[> > >] Condition (3) can not be violated because of similar argument as last case.
                        \item[> > >] Suppose they violate condition (4). Then there is a point $u_1\in W_1$ and $u_1\neq w_1$ such that for any point $u_2\in W_2$ and $u_2\neq w_2$, it is not the case that $(\mathcal{M}_1\backslash\{u_1\}, w_1)\underline{\leftrightarrow}_d (\mathcal{M}_2\backslash\{u_2\}, w_2)$. But again since $n=1$, $u_1\neq w_1$ and $u_2\neq w_2$ can not hold
                        \item[> > >] By similar argument as in previous case, the models cannot violate condition (5).
                    Now we will prove the other side, again by contrapositivity.
                    \item[> >] Conversely, suppose that d-bisimilar(($(\mathcal{M}_1, w_1), (\mathcal{M}_2, w_2), \emptyset)$ returns No, Then one of the following cases occur:
                        \item[> > >] The function returns No at line number 7. This can only happen when the If condition in line 6 is true. Therefore, there exists an atomic proposition $p$ such that, $w_1\in V_1(p)$ but $w_2\not\in V_2(p)$; or $w_1\not\in V_1(p)$ but $w_2\in V_2(p)$. From truth definition of Deletion modal logic, we have either $(\mathcal{M}_1,w_1)\models p$ and $(\mathcal{M}_2,w_2)\not\models p$; or $(\mathcal{M}_1,w_1)\not \models p$ and $(\mathcal{M}_2,w_2)\models p$. But then $(\mathcal{M}_1, w_1)\cancel{$\underline{\leftrightarrow}$}_d (\mathcal{M}_2, w_2)$ as they violate condition (1) of the definition of d-bisimilarity.
                        \item[> > >] The function returns No at line number 16. This can only happen if condition at line 15 is true. But since $|W_1|=1$, $u_1\neq w_1$ can not hold. Hence line 16 is not executed.
                        \item[> > >] By similar argument as previous case, line 24 is not executed and hence function cannot return No from line 25.
                        \item[> > >] Suppose the function returns no from line 37. then the condition at line 36 is true. Therefore $u_2\neq w_2$ is true. But this cannot be the case as $|W_2|=1$.
                        \item[> > >] By similar argument as previous case, the function cannot return No from line 48
                This completes both sides of the base case.
                \item[>] \textbf{Induction Hypothesis:} Suppose the theorem holds good for $n\leq k$. That is, $(\mathcal{M}_1, w_1)\underline{\leftrightarrow}_d (\mathcal{M}_2, w_2)$ iff the function \textsf{d-Bisimilar}($(\mathcal{M}_1, w_1), (\mathcal{M}_2, w_2), \emptyset)$ returns yes when $|W_1|=|W_2|\leq k$
                \item[>] \textbf{Induction Step:} Let $n=k+1$\\
                We will first prove that if $(\mathcal{M}_1, w_1)\underline{\leftrightarrow}_d (\mathcal{M}_2, w_2)$ then the function \textsf{d-Bisimilar}($(\mathcal{M}_1, w_1), (\mathcal{M}_2, w_2), \emptyset)$ returns yes. Again we will prove this by contrapositivity.
                    \item[> >] Suppose the function returns No. Then it executes one of the 6 return No statements. But it can not return NO at line 5, as we have assumed $|W_1|=|W_2|$. So the following cases can occur:
                        \item[> > >] The function returns No at line number 7. This can only happen when the If condition in line 6 is true. But then, by argument similar to that in base case, $(\mathcal{M}_1, w_1)\cancel{$\underline{\leftrightarrow}$}_d (\mathcal{M}_2, w_2)$ as they violate condition (1) of the definition of d-bisimilarity.
                        \item[> > >] The function returns No at line number 16. Then condition in line 15 is true even after execution of forloop at line 10. Therefore, there is some $u_1\in W_1$ and $u_1\neq w_1$ such that for all $u_2\in W_2$ and $u_2\neq w_2$, condition in line 12 is false, $i.e.$ there is an $u_1\in W_1$ and $u_1\neq w_1$ such that for all $u_2\in W_2$ and $u_2\neq w_2$, we have \textsf{d-Bisimilar}($(\mathcal{M}_1\backslash\{u_1\},w_1), (\mathcal{M}_2\backslash\{u_2\},w_2), \emptyset)$ returns NO. But the model $\mathcal{M}_1^\prime=\mathcal{M}_1\backslash\{u_1\}$ and $\mathcal{M}_2^\prime=\mathcal{M}_2\backslash\{u_2\}$ have k points. Therefore, by induction hypothesis, $(\mathcal{M}_1\backslash\{u_1\}, w_1)\cancel{$\underline{\leftrightarrow}$}_d (\mathcal{M}_2\backslash\{u_2\}, w_2)$ for all $u_2\in W_2$ and $u_2\neq w_2$. This is violation to condition (4) in the definition of d-bisimilarity. Therefore, $(\mathcal{M}_1, w_1)\cancel{$\underline{\leftrightarrow}$}_d (\mathcal{M}_2, w_2)$
                        \item[> > >] The function returns no on line 25. By similar argument as in previous case, this leads to violation of condition (5) in the definition of s-bisimilarity. Hence, $(\mathcal{M}_1, w_1)\cancel{$\underline{\leftrightarrow}$}_d (\mathcal{M}_2, w_2)$
                        \item[> > >] The function returns No at line 37. Then condition at line 36 is true for some $u_1\in W_1$. Therefore, following cases arise:
                            \item[> > > >] For a successor $u_1$ of $w_1$, condition at line 30 is false for all $u_2\in W_2$, i.e., $w_2R_2u_2$ is not true for any $u_2\in W_2$. This is a violation of condition (2) in definition of s-bisimilarity and hence $(\mathcal{M}_1, w_1)\cancel{$\underline{ \leftrightarrow}$}_d (\mathcal{M}_2, w_2)$
                            \item[> > > >] Condition at line 31 is true but condition at line 32 is false, i.e., $\exists u_1\in W_1$ such that $w_1R_1u_1$, $\forall u_2\in R_2$ such that $w_2R_2u_2$ and $L$ is such that $(u_1, u_2)\not\in L$ (and $(w_1,w_2)\not\in L$ because line 31 can be executed only if condition in line 26 is true); we get \textsf{d-Bisimilar}($(\mathcal{M}_1, u_1), (\mathcal{M}_2, u_2), L\cup\{(w_1,w_2)\})$) returns No.\\
                            To prove:$(\mathcal{M}_1, w_1)\cancel{$\underline{\leftrightarrow}$}_d (\mathcal{M}_2, w_2)$.\\
                            Proof by induction on $m=|W_1\times W_2|-|L\cup\{(w_1,w_2)\}|$
                                \item[> > > > >] \textbf{Base case:} $|W_1\times W_2|=|L\cup\{(w_1,w_2)\}|$\\
                                We need to prove that if $\exists u_1\in W_1$ such that $w_1R_1u_1$, $\forall u_2\in R_2$ such that $w_2R_2u_2$ and $L$ is such that $(u_1, u_2)\not\in L$ (and $(w_1,w_2)\not\in L$ because line 31 can be executed only if condition in line 26 is true) and $|W_1\times W_2|-|L\cup\{(w_1,w_2)\}|=0$; and \textsf{d-Bisimilar}($(\mathcal{M}_1, u_1), (\mathcal{M}_2, u_2), L\cup\{(w_1,w_2)\})$) returns No, then $(\mathcal{M}_1, w_1)\cancel{$\underline{\leftrightarrow}$}_d (\mathcal{M}_2, w_2)$\\
                                But since $|W_1\times W_2|-|L\cup\{(w_1,w_2)\}|=0$, we have $(u_1,u_2)\in L\cup\{(w_1,w_2)\}$. This is in contradiction with condition in line 31 being true. So the antecedent is false and hence base case is true vacuously.
                                \item[> > > > >] \textbf{Induction Hypothesis:} Suppose the claim holds for $m\leq l$, $i.e.$, \\
                                Suppose whenever $\exists u_1\in W_1$ such that $w_1R_1u_1$, $\forall u_2\in R_2$ such that $w_2R_2u_2$ and $L$ is such that $(u_1, u_2)\not\in L$ (and $(w_1,w_2)\not\in L$ because line 31 can be executed only if condition in line 26 is true) and $|W_1\times W_2|-|L\cup\{(w_1,w_2)\}|\leq l$; and \textsf{d-Bisimilar}($(\mathcal{M}_1, u_1), (\mathcal{M}_2, u_2), L\cup\{(w_1,w_2)\})$) returns No, then $(\mathcal{M}_1, w_1) \cancel{$\underline{\leftrightarrow}$}_d (\mathcal{M}_2, w_2)$
                                \item[> > > > >] \textbf{Induction step:} Suppose $m=l+1$.\\
                                In this case, suppose condition in line 31 is true and condition in line 32 is false. Therefore, we have, $\exists u_1\in W_1$ such that $w_1R_1u_1$, $\forall u_2\in R_2$ such that $w_2R_2u_2$ and $L$ is such that $(u_1, u_2)\not\in L$ (and $(w_1,w_2)\not\in L$ because line 31 can be executed only if condition in line 26 is true) and $|W_1\times W_2|-|L\cup\{(w_1,w_2)\}|= l+1$; and \textsf{d-Bisimilar}($(\mathcal{M}_1, u_1), (\mathcal{M}_2, u_2), L\cup\{(w_1,w_2)\})$) returns No. Now,  \textsf{d-Bisimilar}($(\mathcal{M}_1, u_1), (\mathcal{M}_2, u_2), L\cup\{(w_1,w_2)\})$) can return No either at one of 6 return No statements. If it returns No at lines 4, 7, 16 or 25, then by above cases, we have already proved that $(\mathcal{M}_1, u_1) \cancel{$\underline{\leftrightarrow}$}_d (\mathcal{M}_2, u_2)$ because they violate conditions (1) or (4) or (5) in the definition of d-bisimilarity. Suppose it returns No at line 37, then if condition at line 30 is always false, then $(\mathcal{M}_1, u_1) \cancel{$\underline{\leftrightarrow}$}_d (\mathcal{M}_2, u_2)$ because they violate condition (2) of definition of d-bisimilarity. So suppose condition at line 31 is true but at line 32 is false. Therefore, $\exists v_1\in W_1$ such that $u_1R_1v_1$ and $\forall v_2\in W_2$ such that $u_2R_2v_2$, $L$ is such that $(v_1,v_2)\not\in L\cup\{(w_1,w_2)\}$ (also $(u_1,u_2)\not\in L\cup\{(w_1,w_2)\}$ because condition at line26 has to be true), we have \textsf{d-Bisimilar}($(\mathcal{M}_1,v_1), (\mathcal{M}_2,v_2), L\cup\{(w_1,w_2), (u_1,u_2)\}$) returns No. Now by induction hypothesis, $(\mathcal{M}_1, v_1) \cancel{$\underline{\leftrightarrow}$}_d (\mathcal{M}_2, v_2)$ which implies $(\mathcal{M}_1, u_1) \cancel{$\underline{\leftrightarrow}$}_d (\mathcal{M}_2, u_2)$ and hence $(\mathcal{M}_1, w_1) \cancel{$\underline{\leftrightarrow}$}_d (\mathcal{M}_2, w_2)$.
                        \item[> > >] The function returns No at line 48, then by argument similar to last case, $(\mathcal{M}_1, w_1) \cancel{$\underline{\leftrightarrow}$}_d (\mathcal{M}_2, w_2)$.
                    We will now prove the remaining side by contrapositivity.
                    \item[>] Suppose $(\mathcal{M}_1, w_1)\cancel{$\underline{\leftrightarrow}$}_d (\mathcal{M}_2, w_2)$. Then these models must violate one of the 5 conditions in definition of d-bisimilarity.
                        \item[> >] Suppose they violate condition (1). There there is some atomic proposition $p$ such that either $(\mathcal{M}_1,w_1)\models p$ and $(\mathcal{M}_2,w_2)\not\models p$; or $(\mathcal{M}_1,w_1)\not \models p$ and $(\mathcal{M}_2,w_2)\models p$. From truth definition of deletion modal logic, we have $w_1\in V_1(p)$ but $w_2\not\in V_2(p)$; or $w_1\not\in V_1(p)$ but $w_2\in V_2(p)$. In this case the function returns NO in line 7.
                        \item[> >] Suppose they violate condition (4). Then there is a point $u_1\in W_1$ and $u_1\neq w_1$ such that for any point $u_2\in W_2$ and $u_2\neq w_2$, $(\mathcal{M}_1\backslash\{u_1\}, w_1)\cancel{$\underline{\leftrightarrow}$}_d (\mathcal{M}_2\backslash\{u_2\}, w_2)$. In this case for $u_1$, condition in line 12 is never true by induction hypothesis ($\mathcal{M}_1\backslash\{u_1\}$ and $\mathcal{M}_2\backslash\{u_2\}$ have k edges, hence we can use induction hypothesis). Therefore, return No is executed in line 16.
                        \item[> >] Suppose they violate condition (5), by similar argument as previous case, by induction hypothesis, function returns No.
                        \item[> >] Suppose they violate condition (2) and/or (3). We need to prove if $(\mathcal{M}_1,w_1)\cancel{$\underline{\leftrightarrow}$}_d(\mathcal{M}_2,w_2)$ because they violate condition (2) and/or (3), but not (1), (4) or (5) in the definition of d-bisimilarity, then
                        \textsf{d-bisimilar}(($(\mathcal{M}_1, w_1)$ ,$ (\mathcal{M}_2, w_2), L)$ returns No, for $|L|=0$.\\
                        Since $(\mathcal{M}_1,w_1)\cancel{$\underline{\leftrightarrow}$}_d(\mathcal{M}_2,w_2)$ because they violate condition (2) and/or (3), therefore $\exists u_{11}\in W_1$, $w_1R_1u_{11}$, such that $\forall u_{21}\in W_2$, $w_2R_2u_{21}$, $(\mathcal{M}_1, u_{11})\cancel{$\underline{\leftrightarrow}$}_d (\mathcal{M}_2, u_{21})$ (if condition (2) is violated); or  $\exists u_{12}\in W_2$, $w_2R_2u_{12}$, such that $\forall u_{11}\in W_1$, $w_1R_1u_{11}$, $(\mathcal{M}_1, u_{11})\cancel{$\underline{\leftrightarrow}$}_d (\mathcal{M}_2, u_{21})$. Now if $(\mathcal{M}_1, u_{11})\cancel{$\underline{\leftrightarrow}$}_d (\mathcal{M}_2, u_{21})$ because they violate conditions (1), (4) or (5), then by previous cases, the function returns No at line 7, 16 or 25 respectively and we will be done. Let us pick a general such pair $(v_{11},v{21})$. So, assume $(\mathcal{M}_1, v_{11})\cancel{$\underline{\leftrightarrow}$}_d (\mathcal{M}_2, v_{21})$ because they violate condition(s) (2) and/or (3). Therefore, again, $\exists u_{12}\in W_1$, $v_{11}R_1u_{12}$, such that $\forall u_{22}\in W_2$, $v_{12}R_2u_{22}$, $(\mathcal{M}_1, u_{12})\cancel{$\underline{\leftrightarrow}$}_d (\mathcal{M}_2, u_{22})$ (if they violate (2)); or $\exists u_{22}\in W_2$, $v_{12}R_2u_{22}$, such that $\forall u_{12}\in W_1$, $v_{11}R_1u_{12}$, $(\mathcal{M}_1, u_{12})\cancel{$\underline{\leftrightarrow}$}_d (\mathcal{M}_2, u_{22})$ (if they violate condition (3)). Again, choose a general such pair $(v_{12}, v_{22})$ from above such that $v_{11}R_1v_{12}$ and $v_{21}R_2v_{22}$ and $(\mathcal{M}_1, v_{12})\cancel{$\underline{\leftrightarrow}$}_d (\mathcal{M}_2, v_{22})$. Again, we are done if $(\mathcal{M}_1, v_{12})\cancel{$\underline{\leftrightarrow}$}_d (\mathcal{M}_2, v_{22})$ because they violate condition (1), (4) or (5). So, again, we can assume that they violate condition (2) and /or (3). This can go on until we reach a leaf node, i.e., there is some $k$ such that exactly one of the following is true: $v_{1k}R_1v_{1k+1}$ for some $v_{1k+1}\in W_1$ or $v_{2k}R_2v_{2k+1}$ for some $v_{2k+1}\in W_2$. Again the function returns No, either at line 16 or 25 respectively in both cases. The only case that remains is when there is no leaf nodes and there is some $k$ such that $v_{1k}=v_{1l}$ or $w_1$ and $v_{2k}=v_{2l}$ or $w_2$ for some $l<k$. In this case, since $v_{1i}$ and $v_{2i}$ were some general node in the reachable part from $w_1$ and $w_2$, such that they do not violate condition (1), (4) or (5) in the definition of d-bisimilarity, we have the following:
                            \item[> > >] $(\mathcal{M}_1,w_1)$ and $(\mathcal{M}_2,w_2)$ satisfy conditions (1), (4) and (5) in the definition of d-bisimilarity
                            \item[> > >] For every n, $\exists v_1\in W_1$ such that $w_1R^n_1v_1$ iff $\exists v_2\in W_2$ such that $w_2R^n_2v_2$; and $(\mathcal{M}_1,v_1)$ and $(\mathcal{M}_2,v_2)$ satisfy condition (1), (4) and (5) from the definition of s-bisimilarity.
                        But these conditions are same as the conditions in definition of d-bisimilarity. Hence, $(\mathcal{M}_1, w_1)\underline{\leftrightarrow}_s (\mathcal{M}_2, w_2)$ and the function does not return No in this this case.
                        
            \end{itemize}
     This completes the proof.   
        \end{proof}
        
        \begin{theorem}
            Function d-Bisimilar terminates and is in PSPACE
        \end{theorem}
        \begin{proof}
            The argument is very similar to theorem 10. We look at the depth of the recursion tree. There are four recursive calls in the function \textsf{d-Bisimilar}. Two of them strictly decrease the number of points in the model and other two strictly increase the size of the list. By similar argument to theorem 10, we can bound the depth of the recursion tree by $|W_1|\times |W_1\times W_2|$ which becomes $|W_1|^3$ in this case.
        \end{proof}
        
     \section*{Algorithms for the generalized versions}
        We will now give a similar algorithms to check whether given two pointed models are generalized sabotage bisimilar and generalized point sabotage bisimilar. Algorithms 4 (page 26) and 5 (page 27) decide whether the given pointed models are generalized sabotage bisimilar and generalized point sabotage bisimilar, respectively. The correctness and complexity analyses are similar to theorems 21, 10, 23 and 24. Intuitively, the correctness holds because the algorithms have corresponding changes according to the differences in the definition of various notions of bisimilarity. For complexity, it should be noted that the extra checks do not contribute to the depth of the recursion tree. And since the space complexity depends on the depth of the recursion tree, the following two algorithms remain in PSPACE.\\
        \begin{algorithm}
           \footnotesize
            \LinesNumbered
                \KwIn{ $((W_1,R_1,V_1),w_1), ((W_2,R_2,V_2),w_2)$ } 
                \textbf{Initialize}:L=$\emptyset$\\
                \SetKwFunction{FMain}{s-Bisimilar}
                \SetKwProg{Fn}{Function}{:}{}
                \Fn{\FMain{$((W_1,R_1,V_1),w_1), ((W_2,R_2,V_2),w_2)$, L}}{
                    {\If{$|R_1|\neq |R_2|$}
                    {\Return NO;}
                    }

                    {
                        \ForAll{atomic propositions p}
                        {
                            \If{$(((w_1\in V_1(p)) AND (w_2\not\in V_2(p)))$ OR $((w_1\not\in V_1(p)) AND (w_2\in V_2(p))))$}
                                {
                                    \Return NO;
                                }
                        }
                        \ForAll{$(u_1,v_1)\in R_1$}
                            {
                                Found=0;\\
                                \ForAll{$(u_2,v_2\in R_2$}
                                    {
                                        \If{(\textsf{s-Bisimilar}$(((W_1,R_1,V_1),u_1)$,$((W_2,R_2,V_2),u_2), \emptyset)$==YES) AND (\textsf{s-Bisimilar}$(((W_1,R_1,V_1),v_1)$,$((W_2,R_2,V_2),v_2), \emptyset)$==YES)}
                                        {
                                            \If {\textsf{s-Bisimilar}$(((W_1,R_1\backslash\{(u_1,v_1)\},V_1),w_1)$,$((W_2,R_2\backslash\{(u_2,v_2)\},V_2),w_2), \emptyset)$==YES}
                                            {
                                                Increment Found;\\
                                                break;
                                            }

                                        }
                                                                            }
                                \If{Found =0}
                                    {
                                        \Return No;
                                    }
                            }
                        
                        \ForAll{$(u_2,v_2)\in R_2$}
                            {
                                Found=0;\\
                                \ForAll{$(u_1,v_1)\in R_1$}
                                    {
                                        \If{(\textsf{s-Bisimilar}$(((W_1,R_1,V_1),u_1)$,$((W_2,R_2,V_2),u_2), \emptyset)$==YES) AND (\textsf{s-Bisimilar}$(((W_1,R_1,V_1),v_1)$,$((W_2,R_2,V_2),v_2), \emptyset)$==YES)}
                                        {
                                            \If {\textsf{s-Bisimilar}$(((W_1,R_1\backslash\{(u_1,v_1)\},V_1),w_1)$,$((W_2,R_2\backslash\{(u_2,v_2)\},V_2),w_2), \emptyset)$==YES}
                                            {
                                                Increment Found;\\
                                                break;
                                            }

                                        }
                                    }    
                                \If{Found =0}
                                    {
                                        \Return No;
                                    }
                            }


%

                            \If {$(w_1,w_2)\not\in L$}
                            {
                            \ForAll{$u_1\in W_1$}
                            {
                                Found=0;\\
                                \ForAll{$u_2\in W_2$}
                                    {
                                        \If {($(w_1R_1u_1)$ AND $(w_2R_2u_2))$}
                                        {
                                            \If{$((u_1,u_2)\not\in L)$}
                                            {
                                                \If {\textsf{s-Bisimilar}$(((W_1,R_1,V_1),u_1),((W_2,R_2,V_2),u_2),L\cup\{(w_1,w_2)\})$==YES}
                                                {
                                                    Increment Found;
                                                }
                                            }
                                            \Else{Increment Found}
                                        }
                                            
                                    }
                                \If{(Found=0) AND $(w_1R_1u_1)$}
                                    {
                                        \Return No;
                                    }
                            }
                            
                        \ForAll{$u_2\in W_2$}
                            {
                                Found=0;\\
                                \ForAll{$u_1\in W_1$}
                                    {
                                        \If {($(w_1R_1u_1)$ AND $(w_2R_2u_2))$}
                                        {
                                            \If{$((u_1,u_2)\not\in L)$}
                                            {
                                                \If {\textsf{s-Bisimilar}$(((W_1,R_1,V_1),u_1),((W_2,R_2,V_2),u_2),L\cup\{(w_1,w_2)\})$==YES}
                                                {
                                                    Increment Found;
                                                }
                                            }
                                            \Else{Increment Found}
                                        }
                                            
                                    }
                                \If{(Found=0) AND $(w_2R_2u_2)$}
                                    {
                                        \Return No;
                                    }
                            }
                        }
                        
                        \Return Yes;
                    }
                }
                \caption{Algorithm to check if two models are s-bisimilar in generalized sense}
            \end{algorithm}

        \begin{algorithm}
        \footnotesize
            \LinesNumbered
                \KwIn{ $((W_1,R_1,V_1),w_1), ((W_2,R_2,V_2),w_2)$ } 
                \textbf{Initialize}:L=$\emptyset$\\
                \SetKwFunction{FMain}{d-Bisimilar}
                \SetKwProg{Fn}{Function}{:}{}
                \Fn{\FMain{$((W_1,R_1,V_1),w_1), ((W_2,R_2,V_2),w_2)$, L}}{
                    {\If{$|W_1|\neq |W_2|$}
                    {\Return NO;}
                    }

                    {
                        \ForAll{atomic propositions p}
                        {
                            \If{$(((w_1\in V_1(p)) AND (w_2\not\in V_2(p)))$ OR $((w_1\not\in V_1(p)) AND (w_2\in V_2(p))))$}
                                {
                                    \Return NO;
                                }
                        }
                        \ForAll{$u_1\in W_1$}
                            {
                                Found=0;\\
                                \ForAll{$u_2\in W_2$}
                                    {
                                        \If{$(u_1\neq w_1) AND (u_2\neq w_2)$ AND (\textsf{d-Bisimilar}$((successor((W_1,R_1,V_1),u_1), w_1)$,$(successor((W_2,R_2,V_2),u_2), w_2), \emptyset)$==YES)}
                                        {
                                        \If {\textsf{d-Bisimilar}$((successor((W_1,R_1,V_1),u_1), w_1)$,$(successor((W_2,R_2,V_2),u_2), w_2), \emptyset)$==YES}
                                            {
                                                Increment Found;
                                                break;
                                            }
                                    }
                                    }
                                \If{Found =0 $\wedge (u_1\neq w_1)$}
                                    {
                                        \Return No;
                                    }
                            }
                        
                        \ForAll{$u_2\in W_2$}
                            {
                                Found=0;\\
                                \ForAll{$u_1\in W_1$}
                                    {
                                        \If{$(u_1\neq w_1) AND (u_2\neq w_2)$ AND (\textsf{d-Bisimilar}$((successor((W_1,R_1,V_1),u_1), w_1)$,$(successor((W_2,R_2,V_2),u_2), w_2), \emptyset)$==YES)}
                                        {
                                        \If {\textsf{d-Bisimilar}$((successor((W_1,R_1,V_1),u_1), w_1)$,$(successor((W_2,R_2,V_2),u_2), w_2), \emptyset)$==YES
                                        }
                                            {
                                                Increment Found;
                                                break;
                                            }
                                    }
                                    }
                                \If{Found =0 $\wedge (u_2\neq w_2)$}
                                    {
                                        \Return No;
                                    }
                            }

                            
%

                            \If {$(w_1,w_2)\not\in L$}
                            {
                            \ForAll{$u_1\in W_1$}
                            {
                                Found=0;\\
                                \ForAll{$u_2\in W_2$}
                                    {
                                        \If {($(w_1R_1u_1)$ AND $(w_2R_2u_2))$}
                                        {
                                            \If{$((u_1,u_2)\not\in L)$}
                                            {
                                                \If {\textsf{d-Bisimilar}$(((W_1,R_1,V_1),u_1),((W_2,R_2,V_2),u_2),L\cup\{(w_1,w_2)\})$==YES}
                                                {
                                                    Increment Found;
                                                }
                                            }
                                            \Else{Increment Found}
                                        }
                                            
                                    }
                                \If{(Found=0) AND $(w_1R_1u_1)$}
                                    {
                                        \Return No;
                                    }
                            }
                            
                        \ForAll{$u_2\in W_2$}
                            {
                                Found=0;\\
                                \ForAll{$u_1\in W_1$}
                                    {
                                        \If {($(w_1R_1u_1)$ AND $(w_2R_2u_2))$}
                                        {
                                            \If{$((u_1,u_2)\not\in L)$}
                                            {
                                                \If {\textsf{d-Bisimilar}$(((W_1,R_1,V_1),u_1),((W_2,R_2,V_2),u_2),L\cup\{(w_1,w_2)\})$==YES}
                                                {
                                                    Increment Found;
                                                }
                                            }
                                            \Else{Increment Found}
                                        }
                                            
                                    }
                                \If{(Found=0) AND $(w_2R_2u_2)$}
                                    {
                                        \Return No;
                                    }
                            }
                        }
                        
                        \Return Yes;
                    }
                }
                \caption{Algorithm to check if given models are d-Bisimilar in generalized sense}
            \end{algorithm}

\end{document}